# Electron conductive self-assembled hybrid low-molecular weight glycolipid-nanosilver gels

Korin Gasia Ozkaya[a], Othmane Darouich,[a] Hynd Remita[b], Isabelle Lampre[b], Lionel Porcar[c], Alain Carvalho[d], Marc Schmutz[d], Sandra Casale,[e] Christel Laberty-Robert[a], Niki Baccile[a,*]

[a] *Sorbonne Université, Centre National de la Recherche Scientifique, Laboratoire de Chimie de la Matière Condensée de Paris (LCMCP), UMR 7574, F-75005 Paris, France*

[b] *Université Paris-Saclay, CNRS, Institut de Chimie Physique (ICP), UMR 8000, Faculté des Sciences d'Orsay, 91405 Orsay, France*

[c] *Institut Laue Langevin, 38042 Grenoble, France*

[d] *Université de Strasbourg, CNRS, Institut Charles Sadron UPR 22, 67034 Strasbourg, France*

[e] *Sorbonne Université, Centre National de la Recherche Scientifique, Laboratoire de Réactivité de Surface (LRS), UMR 7574, F-75005 Paris, France*

**Niki Baccile,* niki.baccile@sorbonne-universite.fr

**Abstract**

Low-molecular weight (LMW) hydrogels are gaining interest over macromolecular gels due to their reversible, dynamic and stimuli-responsive nature. They are potentially interesting functional materials for advanced applications such as catalysis, nanoelectronics or regenerative medicine. One common strategy to enhance the functional properties is to incorporate inorganic nanostructures. However, simultaneous control of the gel mechanics, shape and size of the nanostructures and functional properties is challenging. Here, a biobased, double amphiphilic, bolaform, single-glucose lipid (containing glucose and COOH in opposite directions) is able to coordinate silver ions, drive the formation of a self-assembled fibrous hydrogel and, after controlling the reduction time (seconds to hours) of the reduction process ($NaBH_4$, ascorbate, γ-rays), stabilize Ag nanoparticles (NPs) of controlled size (2.8 nm ± 13%). The NPs are spontaneously embedded in the fibers' core following a two-dimensional anisotropic long-range order. Precise control of the reduction parameters (ascorbate) drives the formation of Ag nanowires, possibly due to an anisotropic coalescence process of the nanoparticles. Samples containing Ag nanowires have shown an electronic conductive response, observed with impedance spectroscopy. This works shows the potential of biological amphiphiles to develop under soft conditions (pseudo single step, water, room temperature) advanced hybrid organic/inorganic (O/I) materials with a multiscale structure, order and electron conductivity functionality.

## Introduction

Hydrogels are attractive materials for many potential applications in tissue engineering, biomedicine, pharmaceutics and development of wearable soft sensors due to their mechanical flexibility, biocompatibility and mechanical properties, comparable to the ones of biological tissues[1,2]. Covalently cross-linked synthetic hydrogels, among the most common systems, have however some limitations, such as eco- or cytotoxicity and chemical irreversibility. As a more sustainable alternative, biobased hydrogels based on biopolymers but also on low-molecular weight (LMW) amphiphiles have emerged, where the latter display interesting fast, reversible, self-assembly kinetics.

Low-molecular weight gels (LMWG) are defined as gels conceived from the self-assembly of small molecules into elongated structures through non-covalent interactions[3]. Their dynamic and stimuli-responsive nature make them promising candidates for advanced, functional applications such as controlled drug delivery, catalysis, water remediation, nanostructured electronics, regenerative medicine[4]. To enhance the functionality of the gels, one strategy is to add, or generate *in situ*, inorganic (silica, clays, metal oxides, metal) nanostructures with desired functions,[5–7] so enhance the physical, chemical and biological properties but also stability[8–10] of the resulting hybrid organic/inorganic (O/I) gels.[7] For example, Laponite® nanodisks-peptide amphiphile co-assembled hydrogels[11] were shown bioactive and with good rheological properties, while cellulose-silica nanofiber cryogels combined good mechanical strength and thermal insulation.[12]

Among O/I hybrids, nanoparticle containing gels are noteworthy,[9,10,13] as they can enhance catalytic, magnetic and optical properties.[14–16] Mezzenga et al.[17,18] employed amyloid fibrils to template inorganic nanoparticles to develop hybrid gels and aerogels for catalysis, pressure-sensing[19], water purification[20] and metal quantification[21]. Similarly, Rojas et al. [22] and Ikkala et al.[23] have demonstrated the use of cellulose nanocrystals as templates for metal nanoparticles, which exhibit catalytic activity and plasmonic properties. Metallic nanoparticles can also serve as electrically conductive fillers in the hydrogel network if a continuous percolated network is established.[24] Overall, hybrid conductive gels could be used in further applications in the field of renewable energy as an alternative to carbon-based sustainable materials[25] or as environmentally friendly alternatives to polymer binders in batteries[26].

Despite their success,[9,10] O/I nanoparticle gels hybrids display a heterogeneous distribution of the particles, may the latter be synthesized *in situ* or incorporated in the gel. Homogeneous, anisotropic organization of nanoparticles has so far only been observed in systems that do not form gels. For instance, monodimensional alignment of copper nanoparticles was found on the surface of glycolipid nanotubes after calcination,[27] while a more intriguing two-dimensional (2D), yet anisotropic, alignment of Ag nanoparticles was reported in PhD dissertation[28] involving linoleic acid and silver salts. In another noteworthy study, rod-like tobacco mosaic viruses (TMV) and AuNPs are assembled to form a dispersion of helical superlattice wires with good 2D alignment of nanoparticles. However, the engineering of the system is quite complex, as it involves surface functionalization of the AuNPs to achieve size homogeneity and colloidal stabilization as well as TMV rods as template.[29]

A good control over the anisotropic alignment of the nanoparticles can bring novel functions into the macroscopic system and is critical to improve the optical, electronic, catalytic, magnetic and bio-sensing properties.[30,31] However, as shown above, if anisotropically ordered nanoparticles need advanced material's engineering, nanoparticle-LMWG O/I hybrid gels with controlled particle size, order and function were, to the best of our knowledge, never reported. In the case of conductive gels, electron conductivity is generally not reported for nanoparticle-embedded soft systems. When conductivity is observed, it is either explained by the presence of conductive polymers[32,33] or the nature of the conductivity itself is not clearly specified. For instance, Li et al.[34] have reported conductivity in bacterial cellulose gels containing silver, but the authors do not discriminate between ionic and electronic conductivity, an issue often found in the literature.[35,36] Nanowires (NW)-containing gels can better promote conductive networks due to their anisotropic shape. However, most of the hybrid NW gels found in the literature are polymeric networks,[35,36] and if they are supramolecular gels, conductive behavior is not mentioned.[37,38] Moreover, material's processing, like dielectrophoresis[39] or coating,[40] may be needed to generate conductivity.

Here, we show that hybrid O/I gels with electron conductive properties can be prepared in a nearly single step process using a biobased double amphiphilic, bolaform, glycolipid (containing glucose and COOH in opposite directions) in the presence of a source of $Ag^+$ and a reducing agent, in water and at room temperature. Reducing the silver-coordinated biobased glycolipid hydrogel[41,42] generates a dense three-dimensional network of anisotropic fibers, which embed a bidimensional array of Ag NPs of 2.8 nm and narrow (±13%) size distribution. This particular monounsaturated single-glucose lipid, G-C18:1 plays a multiple role: 1) its free carboxylate group coordinates $Ag^+$ ions[43] but it also acts as a capping agent for the silver NPs after reduction; 2) its double amphiphilic structure drives the fibrous gel network.[43]

Control of the reduction time (second to hours), performed using well-known strategies ($NaBH_4$, ascorbate, γ-rays)[44,45] eventually leads to a morphological evolution of the nanoparticles into nanowires, which is responsible for the electrical conductivity of the gel measured by rheo-impedance spectroscopy. Additional advanced structural characterization (simultaneous neutron and X-rays scattering, cryogenic transmission and scanning electron microscopy) help better understanding the structure/properties relationship of the gels across scales, from 2D to 3D.

## Material and methods

*Chemicals.* The single-glucose lipid G-C18:1 (Mw = 460 g $mol^{-1}$, Figure 1), which consists of a β-D-glucose group covalently linked to the ω-1 methylene of oleic acid, is obtained from the fermentation of the genetically-modified yeast *S. bombicola (ΔugtB1)*. Its synthesis and use were reported in previous publications[46]. The same source of the compound is used in this work. The compound was purchased from Amphistar, Belgium. $AgNO_3$ and $AgClO_4$ were purchased from Sigma Aldrich. $NaBH_4$ is obtained from ThermoFisher and L-Ascorbic acid from Sigma Aldrich. NaOH (pellets) is purchased from Sigma Aldrich and HCl (liquid 35 wt%) from VWR.

*Hydrogel Preparation.* Hydrogels were prepared according to a published protocol[41]. G-C18:1 is prepared in a micellar phase in Milli-Q water at pH 8. The pH is adjusted by addition of NaOH and HCl solutions of 5, 1, 0.5 and 0.1 M. $AgNO_3$ and $AgClO_4$ solutions are prepared at 1 M, and 42 μL of either one of the $Ag^+$ solutions are added to the G-C18:1 pH 8 medium ({$Ag^+$}G-C18:1), to make a total of 1 mL at 2 wt% G-C18:1 with cation-to-G-C18:1 ratio of 1.0, which corresponds to a [$Ag^+$]:[$COO^-$] ratio of 1.0 (G-C18:1 is deprotonated at pH 8)[47]. After the metal salt addition, the solution is vortexed for about 15 s. All samples studied here are protected from light at any moment of preparation until analysis to prevent the photochemical reduction of $Ag^+$ ions.

*Reduction of Ag.* The chemical reduction of $Ag^+$ into $Ag^0$ is performed using three different chemical or physical methods (Table 1), known to have an impact on the reduction time.[44,45]
1) Fast (seconds): use of 10 μL of a NaBH in water solution prepared at 300 mg/mL (8 M), that is 0.08M in the gel; 2) Medium (minutes): use of 10 μL of an ascorbic acid (AA) aqueous solution prepared at pH 10 and 300 mg/mL (1.7 M), that is 0.017M in the gel. Either the $NaBH_4$ or the AA solutions are added to 1 mL of the {$Ag^+$}G-C18:1 hydrogels, vortexed and stirred immediately afterwards. The addition of the reducing agents is systematically performed in an ice bath, to slow down the reaction kinetics. After 1 hour on ice, gels are stored in dark. 3) Slow (hours): radiolytic reduction of $Ag^+$ using a $\gamma$-irradiation source with a panoramic $^{60}Co$ $\gamma$-facility, maximum dose rate 2.0 $kGy.h^{-1}$ at the Institut de Chimie Physique, Orsay, France. Radiolysis is a very powerful method to synthesize metal nanoparticles in solutions and in complex media[48]. Here, $Ag^+$ ions are homogeneously reduced by solvated electrons and reducing radicals induced by solvent radiolysis[49]. Gels containing silver cations are inserted in brown glass vials, they are bubbled with nitrogen gas for 5 minutes prior to irradiation, to avoid the oxidation of the atoms produced and then irradiated. The irradiation time is 4h30 (dose 9 kGy) in the present work, while results collected at longer irradiation times will be presented in a further communication. The longest irradiation for 28 h, is a liquid without the gel organization and used as a control for some characterization tools.

**Table 1- Reduction methods of the hydrogels prepared at a concentration of 20 mg/mL. The sample names will be referenced throughout the text.**

| *Sample Name* | *$C_{G-C18:1}$(mgmL$^{-1}$)* | *Reduction method* |
|---|---|---|
| {$Ag^+$}G-C18:1 | 20 | no reduction |
| {$Ag^0$}G-C18:1-$NaBH_4$ | 20 | 10 µL $NaBH_4$ (C= 300 mgmL$^{-1}$) in 1mL {$Ag^+$}G-C18:1 |
| {$Ag^0$}G-C18:1-AA | 20 | 10 µL ascorbic acid (C= 300 mgmL$^{-1}$, pH=10) in 1mL {$Ag^+$}G-C18:1 |
| {$Ag^0$}G-C18:1-radiolysis | 20 | 9 kGy $\gamma$-irradiation |

*Oscillatory rheology.* Measurements are carried out with an MCR 302 rheometer (Anton Paar, Graz, Austria), with a plate-plate geometry (sandblasted, Ø: 25 mm) and an interplate spacing of 1.0 mm. About 0.5 mL of gel is loaded onto the plate and the samples are protected from evaporation with a solvent trap. After sample loading, a stability measurement is employed for 5 minutes at constant strain and frequency before each measurement. The linear viscoelastic regime (LVER) is defined by a strain sweep over the range of 0.001% to 1000% at a frequency of 1 Hz and the frequency sweep experiments are performed over the range of 0.01 to 100 Hz at a strain of 0.01 %.

*Small Angle X-ray Scattering (SAXS).* SAXS measurements were performed with a XEUSS 1.0 laboratory instrument from Xenocs, equipped with a GENIX 3D Cu X-ray source (λ= 1.54 Å). Data are collected on a DECTRIS PILATUS-300K detector. For each sample, two acquisitions are made at the sample-to-detector distance of 0.3 m and 2.3 m. Gels are manually injected into a home-made flow-through capillary (2 mm in diameter borosilicate glass capillary) cell using a 1.0 mL syringe. In between each sample, the capillary is washed with ethanol and water. Data is treated using the SasView 5.0.5 software. Background subtraction is performed in the usual way, by subtracting the signal of milliQ water recorded in the same capillary. All samples, water included, are irradiated exactly at the same position of the capillary.

*SAXS-coupled to small angle neutron scattering (SANS).* Coupled SAXS-SANS experiments were performed at the D22 beamline of the Institut Laue Langevin (ILL, Grenoble, France, EASY-1216 proposal) following a setup, of which the details are given elsewhere. [50] Specifically for the present experiment, the sample-to-detector distances were 8 m and 1.4 m for SANS and 0.56 m for SAXS, and the experiments were performed at a wavelength of 6 Å for SANS and 1.54 Å for SAXS.

*Cryo-TEM.* Cryo-TEM images were recorded on two instruments, a JEOL JEM-2010 operating at 200-kV equipped with a Gatan Ultrascan 4K CCD camera (Laboratoire de Physique du solide, Université Paris Saclay, Saclay, France) and an FEI Tecnai 120 twin microscope operating at 120-kV with an Orius 1000 CCD camera (Laboratoire de Chimie de la Matière Condensée de Paris, Paris, France). In both cases, a Gatan 626 Cryo holder is used as a sample holder. Gels are diluted 10 times in Milli-Q water and a few microliters of sample solution are deposited on a holey carbon coated TEM nickel grid (Quantifoil R2/2, Germany), ionized by glow-discharge. Excess solution is soaked up with a filter paper right before plunging the grid into liquid ethane, and then the grid is transferred into liquid nitrogen, with minimal air contact. During all experiments, the grids are kept at liquid nitrogen temperature, and the acquisitions are made at low dose mode.

*Scanning transmission electron microscopy (STEM) and high resolution transmission electron microscopy (HR-TEM).* STEM, HR-TEM and EDX elemental maps characterizations were performed on a JEOL 2100Plus UHR (LaB6 gun) operating at 200kV equipped with a Gatan Orius 4K CCD camera and an SDD Oxford EDX detector X-max 80mm2 with Aztec

Software (Fédération de Chimie et Matériaux Paris-Centre, FCMAT FR2482, Paris, France). Samples were put on Copper TEM grids and holder is made of gold.

*UV-Visible Spectroscopy.* UV-Vis spectra are collected with a Cary 5000 UV-Vis-NIR spectrophotometer (Agilent Technologies). Gels are diluted in water and examined in 1.5 mL cuvettes with an optical path of 1 cm.

*Fourier transform infrared spectroscopy (FT-IR).* FTIR measurements were carried out with a Perkin Elmer 400 Spectrometer on freeze-dried samples over the range of 4000-500 $cm^{-1}$. Gel samples are dried in Labconco's Freezone Triad Freeze Dryer during 24 hours and then a small amount of powder is put on the optic window with a diamond crystal for FTIR measurement.

*Scanning electron microscopy under cryogenic conditions (cryo- SEM).* A small volume of the $\{Ag^0\}$G-C18:1-AA gel is plunged into liquid ethane and then placed under a high vacuum (10-6 mbar) and low temperature (-150 °C) into the cryo-preparation chamber (Quorum PT 3010) attached to the microscope. There, the frozen sample is fractured with a razor blade and a slight etching of the sample is performed at -90 °C for 3 min. No further metallization step is required before transferring the sample to the SEM chamber. Eventually, the sample was transferred to the FEG-cryo-SEM (Hitachi SU8010) and observed at 1 kV at  150 °C.

*Conductivity measurements.* The conductive behaviour of the gels after reduction is measured by a rheo-impedance spectroscopy set up. An Anton Paar MCR 302 rheometer is used with hood and plate with Peltier heating/cooling (H-PTD200 and P-PTD200) and dielectric measuring accessories. As a measuring system, PP25/DI/TI with insulation is used. The upper geometry (plate-plate, Ø: 25 mm) and lower plate surface act as electrodes for the impedance measurements. The plate and geometry must be carefully cleaned before use, to remove any possible deposit from previous use and which would reduce the conductivity. Gels are poured between two electrodes in order to cover the entire electrode surface and the measuring distance is set as 0.5 mm. To remove the mechanical history of the sample, a constant shear at 1000 $s^{-1}$ is applied during 5 minutes followed by 5 minutes at rest. Impedance measurement is done with a Solartron SI 1260 Impedance Analyzer. Diagrams are recorded between $10^7$ (ionic conductivity domain) and $10^{-3}$ (electron conductivity domain) Hz with an amplitude of 100 mV (acquisition time is 1h30). In the ionic conductivity domain, all ionic species ($H^+$, $OH^-$, residual $Ag^+$, ascorbate, $Na^+$) are measured simultaneously. The value of the amplitude and frequency range are chosen with a double purpose, to observe both the ionic and electron conductivity domains in the same experiment and keep the experimental time reasonable, so to avoid water evaporation effects.

## Results and discussion

It was previously shown that G-C18:1 (Figure 1) forms self-assembled fibrillar network (SAFiN)[41,42,47] or wormlike[41,47] hydrogels depending on the nature of the cation added to its micellar phase at pH> 7. In the specific case of silver, $Ag^+$ ions drive a SAFiN gel with “fishnet”-like structure, unusual for SAFiNs: the silver-complexed G-C18:1 fibers (cross-section of about 5 nm) align as infinitely long, crystalline, 1D stacks with side-by-side lamellar ordering,[41,42] reminiscent of raft-like structures as known for imogolite nanotube structures[51]. The possible gel structure is illustrated in Figure 1. Here, we attempt to reduce the $Ag^+$ cations, holding the fibrous network together via a complexation mechanism of carboxylate groups, into metallic silver, $Ag^0$, using different chemical and physical methods, known to have an impact on the reduction time[44,45]: fast (~seconds) is promoted with $NaBH_4$, slow (~hours) with γ-rays and intermediate (~minutes) with ascorbic acid.

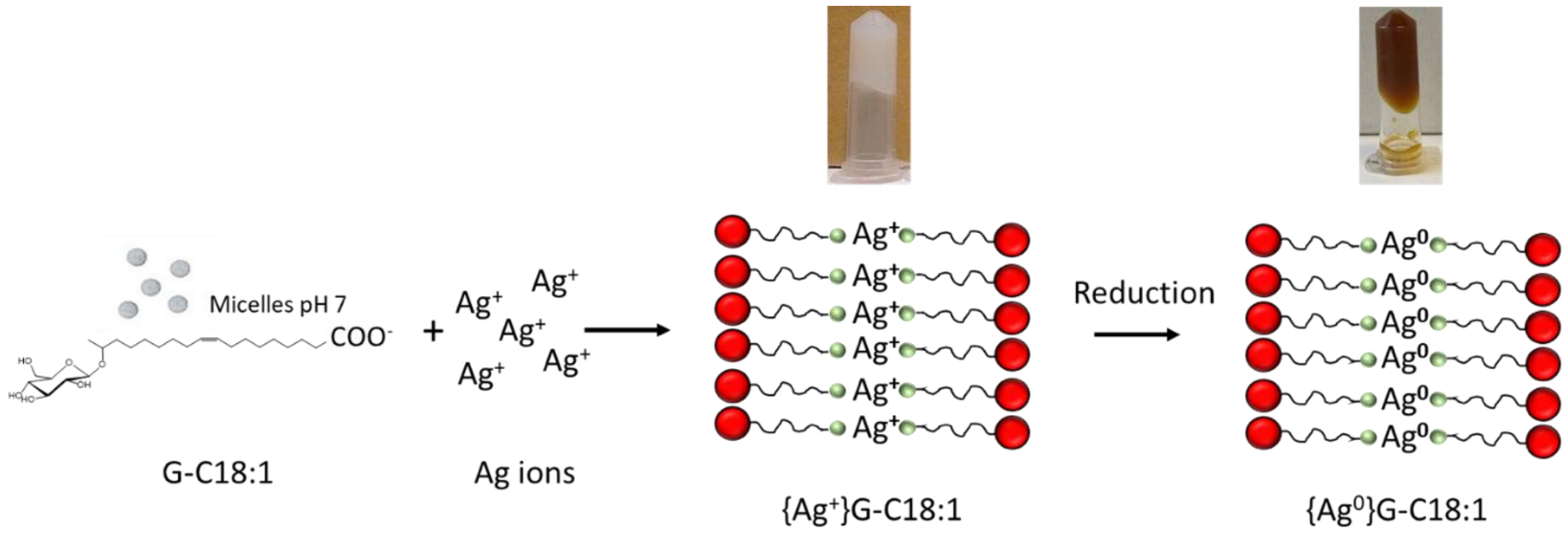

**Figure 1 - Single-glucose lipid G-C18:1 and its $Ag^+$-induced assembly into a hydrogel and the strategy to obtain a silver, $Ag^0$, gel. The possible structure of the {$Ag^+$}G-C18:1 hydrogel was discussed elsewhere.**[42]

### *Anisotropically 2D organized nanoparticles*

The as-prepared, control, {$Ag^+$}G-C18:1 gel is whitish and translucid (Figure 2.a). After reduction, the gel becomes brown/black, indicative of $Ag^0$ (Figure 2.a). By the tube inversion method, only{$Ag^0$}G-C18:1-AA and {$Ag^0$}G-C18:1-radiolysis samples keep the gel state, while {$Ag^0$}G-C18:1-$NaBH_4$ turns into a liquid solution (Figure 2.a). Such visual observation is confirmed by oscillatory rheology performed in the LVER (γ = 0.01%) 24 h after the sample preparation (control or reduced, Figure 2.b). All samples, except {$Ag^0$}G-C18:1-$NaBH_4$, have G' values one order of magnitude greater than G'', indicating the typical elastic behaviour of a gel in the 0.01 Hz to 10 Hz interval. Interestingly, the G' values after reduction are either strictly the same (ascorbic acid) or in the same order of magnitude (radiolysis, 50-80 Pa) as for the control, thus showing that the gels' elastic properties are essentially unchanged after the reduction of silver.

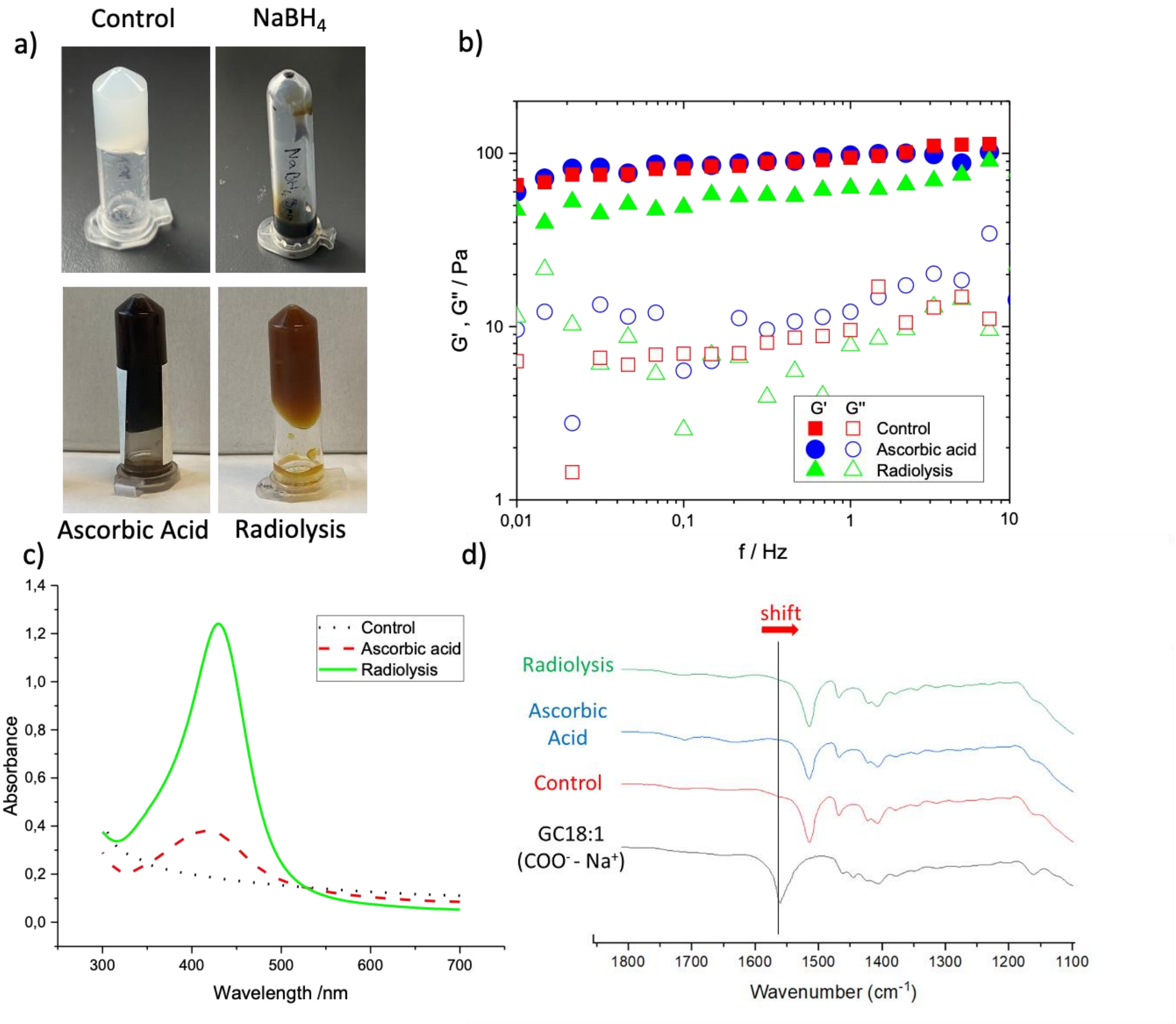

**Figure 2 – Combination of several analytical methods to study reduced {$Ag^0$}G-C18:1 gels. a) tube inversion test, b) oscillatory rheology, c) UV-Vis Spectroscopy on diluted (10x) samples and d) FT-IR Spectroscopy on freeze-dried samples. The sample name control corresponds to {$Ag^+$}G-C18:1, ascorbic acid to {$Ag^0$}G-C18:1-AA, $NaBH_4$ to {$Ag^0$}G-C18:1-$NaBH_4$ and radiolysis to to {$Ag^0$}G-C18:1-radiolysis. Refer to Table 1 for composition of samples.**

Combination of UV-Vis (gel) and FT-IR (freeze-dried) spectroscopies (Figure 2.c,d) confirms the reduction process and shows that the chemical nature of G-C18:1 is not affected by the reduction and suggests a mode of interaction between G-C18:1 and silver. {$Ag^+$}G-C18:1 does not show any UV-Vis absorption band, as expected for G-C18:1 alone. The corresponding full FTIR spectra presented in Figure S 1 in the Supporting Information (SI) show that no major spectroscopic changes occur between the control and the reduced samples (more details given in the SI), thus indicating that the reduction process does not affect the chemical structure of G-C18:1. This being said, the FT-IR experiment shows a shift from 1560 $cm^{-1}$ (asymmetric C=O($COO^-$) stretch) measured for the Ag-free G-C18:1 ($COO^-$ -$Na^+$) sample, prior to silver addition, to 1510 $cm^{-1}$ in the {$Ag^+$}G-C18:1 control gel (Figure 2.d), supporting the formation of silver carboxylate complexes[52,53]. Upon reduction by ascorbic acid or radiolysis, an intense UV-Vis absorption band appears at 420-430 nm, corresponding to the plasmon of Ag nanoparticles[54]. Interestingly, the FTIR peak of $Ag^+$ coordinated carboxylate after reduction is still at 1510 $cm^{-1}$, thus indicating a similar interaction mechanism between the carboxylate group of G-C18:1 and silver metal.

Even if the change in the gel's color and the maintenance of the mechanical properties indicate that the radiolysis and ascorbic acid reduction methods generate silver nanostructures in a gel matrix, it is still unclear whether or not silver is part of the network and what is the morphology and size of silver. Answering this question is crucial, as previous studies have often shown a heterogeneous distribution of polydisperse nanoparticles within hydrogels or they tend to accumulate outside of the gel fibers rather than being incorporated into the fiber network[9,10]. The structural changes by reducing $Ag^+$ into $Ag^0$ are then studied with a combination of cryo-TEM, UV-Vis, and FTIR spectroscopies as well as coupled SAXS-SANS, whereas the use of neutrons is particularly needed to avoid *in situ* reduction of silver by X-rays, actually used as a source of reduction[55–57].

Figure 3 shows the cryo-TEM images of hydrated {$Ag^0$}G-C18:1-AA and {$Ag^0$}G-C18:1-radiolysis (diluted) hydrogels. First of all, the fibrous structure is preserved (Figure 3a,d), while a closer look at each ribbon (Figure 3b,c,e,f,g,h) shows that they are constituted by an ordered array of nanoparticles, keeping the lateral fibers' association preserved. High-magnification HRTEM images recorded on a dried {$Ag^0$}G-C18:1-radiolysis sample (Figure S 2) show the crystalline nature of the Ag nanoparticles, of which the chemical composition (EDX spectra) in Figure S 2c) and typical (111) d-spacing (theoretical, d= 0.21 nm, measured, d= 0.22 nm) (Figure S 2b) confirm the assumption of $Ag^0$.[58]

Evaluation of the NPs size distribution (in %) within the ribbons is attempted by three different methods, two manuals and one automated, as detailed in the Supporting Information, Figure S 3. However, due to the resolution limits, weak contrast and shape variability, the automated method (22%, Figure S 3a) was not considered to be reliable, just as the manual measure of the diameters (16%, Figure S 3b). Therefore, a manual measure of the particles' circumference (> 100 particles) was performed (Figure S 3c), providing NPs of a size of ±2.8 nm with a polydispersity of 13%, as reported in Figure 3i. To our knowledge, such a control over both the size of the nanoparticles and their anisotropic organization within hydrogel fibers has never been reported for metal nanoparticle-embedded hydrogels[59–62]. In most cases, either gelator is not capable to control the nanoparticle size and dispersity, or amphiphilic molecules, that can stabilize nanoparticles, lack gelator properties, as seen in the case of linoleic acid with silver salts[28].

Finally, $\{Ag^0\}$G-C18:1-$NaBH_4$ is only composed of polydisperse, disordered, Ag NPs, as shown in Figure S 4. Also, any attempt to record a meaningful cryo-TEM image for the $\{Ag^+\}$G-C18:1 control was vain due to the *in situ* reduction of $Ag^+$ by electrons, as expected[55,56].

The long-range ordered anisotropic bidimensional organization of the nanoparticles along the ribbons after slow and intermediate reduction time and the loss of such organization after fast reduction ($NaBH_4$) is confirmed by SAXS (Figure 4). Both the control $\{Ag^+\}$G-C18:1 and reduced $\{Ag^0\}$G-C18:1-AA and $\{Ag^0\}$G-C18:1-radiolysis gels exhibit identical SAXS profiles, thus suggesting consistent internal structures. On the other hand, the $\{Ag^0\}$G-C18:1-$NaBH_4$ sample (liquid brown suspension) displays a completely different SAXS pattern, characteristics of individual spherical objects, as similar profiles are found for such objects elsewhere[63–65] , and thus explaining the loss in the elastic properties.

The SAXS pattern of $\{Ag^+\}$G-C18:1, $\{Ag^0\}$G-C18:1-AA and $\{Ag^0\}$G-C18:1-radiolysis (Figure 4b) is similar to previously reported work[42]. Briefly, it is characterized by four broad diffraction peaks at a ratio in wave vector, q, of 1:2:3:4 (with notation $1\vec{x}$, $2\vec{x}$, $3\vec{x}$, $4\vec{x}$, explained later in the paragraph), identifying the side-by-side raft-like association of the fibers, with a typical (001) d-spacing of 51.1 Å. This same structural attribution is coherent with the structural data extracted from fast Fourier transform (FFT) analysis of the corresponding cryo-TEM images (Figure 4a): the side-by-side distance between the ribbon's longest axis (z-axis), along which the $Ag^0$ NPs self-organize after *in situ* reduction, agree well with the (001), (002) and (003) reflections in the corresponding SAXS profiles. The anisotropic organization also facilitates the clear visualization of defects in the self-assembly of G-C18:1, as indicated by the arrows a and b in Figure 4a. The q-values, calculated according to $q = 2\pi/d$, associated to the diffraction spots observed on the FFT (inter-particle distances along the $\vec{x}$ and $\vec{z}$ -axis) nicely match the q-values of the corresponding SAXS peaks (Figure 4.b, Table 2).

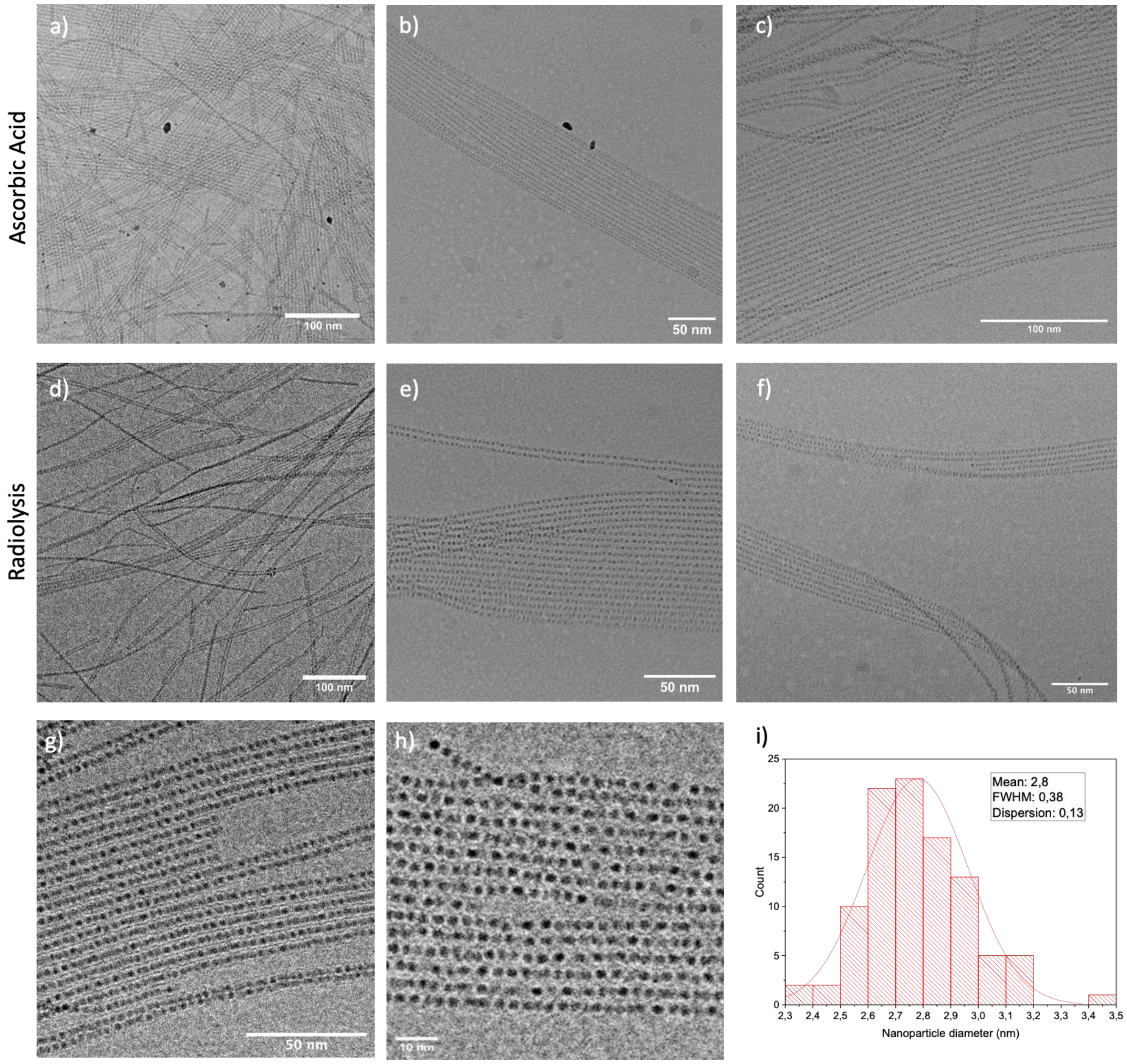

**Figure 3 - Cryo-TEM images of a-c) {Ag$^0$}G-C18:1-AA gel and d-h) {Ag$^0$}G-C18:1-Radiolysis gel. Magnified images (b,c,e,f,g,h) show the anisotropic organization of the nanoparticles along the G-C18:1 fibers. i) Size distribution of nanoparticles (N= 100) is performed by manual count, as reported in Figure S 3c,f. Figure S 3 provides more methods employed in the analysis of the particle size distribution.**

The q-values obtained from the diffraction spots along $\vec{x}$ (see also, Figure S 5b,c) of the FFT pattern (Figure 4.a), q(1$\vec{x}$)= 0.12 Å$^{-1}$, q(2$\vec{x}$)= 0.24 Å$^{-1}$, q(3$\vec{x}$)= 0.36 Å$^{-1}$, are to be compared with q(1$\vec{x}$)= 0.123 Å$^{-1}$, q(2$\vec{x}$)= 0.248 Å$^{-1}$, q(3$\vec{x}$)= 0.367 Å$^{-1}$ and q(4$\vec{x}$)= 0.492 Å$^{-1}$ measured from SAXS, where 1$\vec{x}$, 2$\vec{x}$, 3$\vec{x}$ and 4$\vec{x}$ refer to the (001), (002), (003) and (004) planes associated to the side-by-side lamellar order of the raft-like fibrous structures. The anomalous broader (001) peak seems to be an artefact generated by the strong low-q scattering, as shown in the Kratky representation (Figure S 6).

An interval of q-values (q($1\vec{z}_{min}$)= 0.15 Å$^{-1}$ ; q($1\vec{z}_{max}$)= 0.25 Å$^{-1}$) is observed along $\vec{z}$ of the FFT pattern (see also, Figure S 5b,c) due to a more heterogenous distribution of inter-particle distances along this axis. This q interval corresponds to the area shown in red on the SAXS profile in Figure 4.b, and it most likely explains the stronger intensity of the (002) peak by the superposition of $1\vec{z}$ and $2\vec{x}$. This result can be nicely reproduced by comparing the radial (360°, $\vec{x}$-$\vec{z}$) and arc (60°, $\vec{z}$) integrals and linear ($\vec{x}$) profile of several FFT images performed on a typical cryo-TEM picture (Figure S 5a-c).

Besides demonstrating the origin of the diffraction peaks in SAXS, the combination of SAXS and cryo-TEM also shows that the G-C18:1 homogeneously assembles and surrounds the Ag nanoparticles, which would otherwise cause a shift in the side-by-side inter-fiber distance along the x-axis[66].

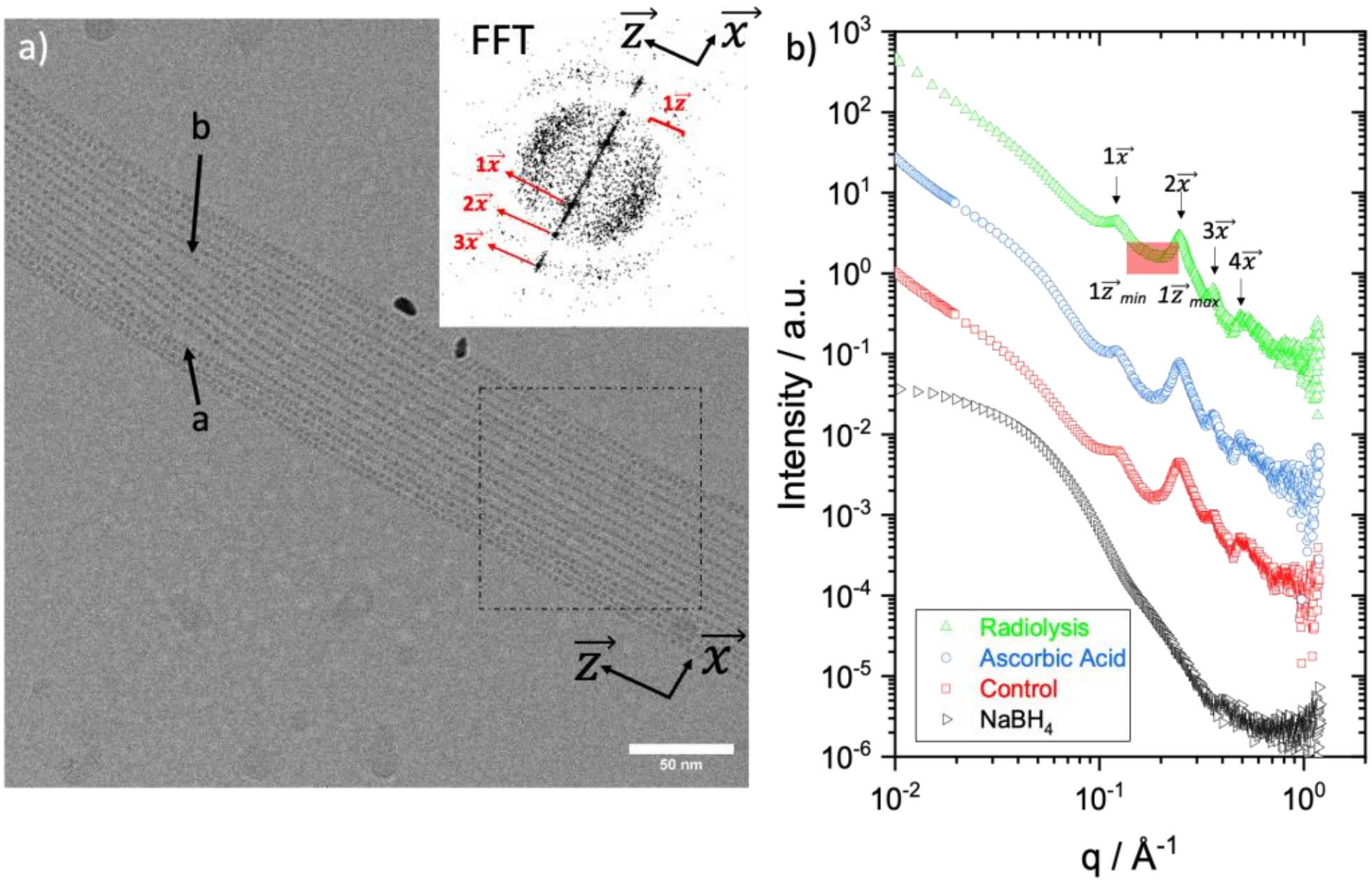

**Figure 4 – a) Cryo-TEM image of {Ag$^0$}G-C18:1-AA, $\vec{x}$ and $\vec{z}$ directions are arbitrarily chosen. The inset shows the fast Fourier transform (FFT) of the segmented square. Arrows a and b indicate defects within the structure, clearly as seen through the anisotropic arrangement. Additional cryo-TEM and FFT are provided in Figure S 5. b) SAXS profiles of the liquid {Ag$^0$}G-C18:1- NaBH$_4$ sample, and gel state {Ag$^+$}G-C18:1 (Control), reduced {Ag$^0$}G-C18:1-AA (Ascorbic acid) and {Ag$^0$}G-C18:1-radiolysis (Radiolysis) samples. Values of $1\vec{x}$, $2\vec{x}$, $3\vec{x}$ and $1\vec{z}$ are given in Table 2.**

A closer look at Figure 4.b shows that the SAXS pattern of the control {Ag$^+$}G-C18:1 displays the same profile as found in the reduced (radiolysis, ascorbic acid) samples. If this can be explained by the *in situ* reduction into Ag$^0$ by X-rays, as observed in similar systems[55,56], further experiments, coupling SAXS and SANS, were attempted to better understand their unbiased structures. Neutrons are complementary to X-rays to study the structure of matter due to the differences in its scattering length densities (sld) towards these sources of radiation[67].

Indeed, the contrast between silver and water ($\Delta$(sld)= 68.4 $\cdot 10^{-6}$ Å$^{-2}$) using X-rays is so high that the spatial organization of the silver nanoparticles dominates the SAXS scattered intensity both on the control and reduced (gel) sample (Figure 5a-SAXS illustrations). On the other hand, when a neutron source is used, the contrast between silver and (heavy) water is much smaller ($\Delta$(sld)= 2.9 $\cdot 10^{-6}$ Å$^{-2}$), and actually smaller compared to the contrast between organic matter and $D_2O$ ($\Delta$(sld) ~6 $\cdot 10^{-6}$ Å$^{-2}$)

**Table 2 –Attribution of the diffraction spots on the FFT pattern corresponding to cryo-TEM and to the q-values of the SAXS peaks.**

| | | SAXS | | Cryo-TEM | |
|---|---|---|---|---|---|
| Annotation | Attribution | $q_{\vec{x}}$ / Å$^{-1}$ | $d_{\vec{x}}$ / Å | $d_{\vec{x}}$/ Å | $q_{\vec{x}}$ / Å$^{-1}$ |
| $1\vec{x}$ | $(001)_{\vec{x}}$ | **0.123** | 51.1 | **51.5** | 0.12 |
| $2\vec{x}$ | $(002)_{\vec{x}}$ | **0.248** | 25.3 | **25.9** | 0.24 |
| $3\vec{x}$ | $(003)_{\vec{x}}$ | **0.367** | 17.1 | **17.3** | 0.36 |
| $4\vec{x}$ | $(004)_{\vec{x}}$ | **0.492** | 12.8 | | |
| | | SAXS | | Cryo-TEM | |
| Annotation | Attribution | $q_{\vec{z}}$ / Å$^{-1}$ | $d_{\vec{z}}$ / Å | $d_{\vec{z}}$ / Å | $q_{\vec{z}}$ / Å$^{-1}$ |
| $1\vec{z}_{min}$ | $(001)_{\vec{z}\,min}$ | **0.153** | 41.1 | **39.5** | 0.15 |
| $1z_{max}$ | $(001)_{\vec{z}\,max}$ | **0.216** | 29 | **25.1** | 0.25 |

Another advantage of neutrons is that they do not reduce silver, even though one could question the reduction by gamma rays. Although possible, this event is most likely excluded, as the acquisition time for a single SANS spectrum is 5 min, while $\gamma$-rays reduction experiments have shown that at least 2 hours of continuous irradiation are required to start to observe a change in color (from white to yellow-brown). For these reasons, SANS is essentially sensitive to the fiber-solvent contrast both in the control and in the reduced samples (Figure 5a-SANS illustrations). Finally, in order to differentiate the scattering signals of fibers and nanoparticles, the liquid sample irradiated by $\gamma$-rays for 28 hours is used as a control. Indeed, in this sample, nanoparticles and fibers coexist, but without co-alignement. Therefore, SAXS characterizes a suspension of nanoparticles (Figure 5a-SAXS/Liquid illustration), while SANS a dispersion of fibers (Figure 5a-SANS/Liquid illustration).

A unique simultaneous SAXS-SANS (D22 beamline, ILL) coupled experiment (Figure 5b) was then performed on the control {$Ag^+$}G-C18:1 and reduced {$Ag^0$}G-C18:1 (gel and liquid) samples in order to better understand both the impact of the silver nanoparticles on the fibers' organization and the fibers' structure prior to reduction. First of all, the scattering profiles of the overly reduced liquid {$Ag^0$}G-C18:1-radiolysis (28h) sample are completely different if probed by SAXS (7 in Figure 5b) or SANS (6 in Figure 5b), performed on the same sample, where both techniques are combined. The former signal is representative of spherical objects [63–65] while the latter is characteristic of anisotropic flat structures (slope of -1.8 at low-q)[47]. This control experiment demonstrates that in the absence of co-alignment, the SAXS and SANS profiles are different for the same sample. Similarly, combined SAXS and SANS profiles of the same {$Ag^+$}G-C18:1 control sample (1 and 2 in Figure 5b) are also quite different. The SAXS pattern, dominated by the $Ag^0$ NPs (*in situ* reduction), reveals the long-range side-by-side arrangement of the fibers ($1\vec{x}$, $2\vec{x}$). In the SANS spectrum, on the contrary, one recognizes the profile of individual fibers, same as for 6 in Figure 5b, but with a contribution at 0.12 $Å^{-1}$ ($1\vec{x}$) and 0.24 $Å^{-1}$ ($2\vec{x}$), identified as a lamellar periodicity due to the raft.like association of the fibers. Indeed, very similar profiles are observed for all reduced samples (3, 4, 5 in Figure 5b), except for 6 in Figure 5b, the only sample in which fibers do not align. Interestingly, neither the diffraction peaks of SAXS, nor the form factor of SANS are shifted upon reduction, thus indicating that the NPs are well embedded in the network of G-C18:1 ribbons and their presence neither disrupt the organization of the fibers nor cause significant displacement of the glycolipids within the fibers themselves.

The long-range order of uniform nanoparticles with controlled size is indeed impressive and challenging to attain in gel networks, because nanoparticles tend to aggregate in the absence of a capping agent. Results demonstrating such a long-range order in the literature are rare, as the nanoparticles' size and their distribution in the gel is rarely controlled.[9,10] In situ reduction of metal ions associated to LMWGs has been employed a number of times in the literature to generate O/I hybrids, nanoparticle containing gels. Typically, bile salts[68,69], peptides[70], bipyridyl-based[71], peptide amphiphiles[72,73], fluorenylmethoxycarbonyl (FMOC)-based gelators[74–76] and even combination of FMOC and peptides[44,77] are among the most studied systems, recently reviewed by Cametti et al.[9] If all approaches generate nanoparticles-containing gels, challenges are still high. Size distribution is often difficult to achieve, with few exceptions[69,78]; volumetric distribution of the NPs is rarely controlled[78]; long-range ordered alignment of the NPs with the gelator's fibers is generally not obtained. Few spurious exceptions should be noted. Within the domain of hydrogels, Li et al. [69] could control the growth of individual Ag nanoparticles at spatially arranged locations along sodium deoxycholate nanohelices by UV reduction of silver ions. Out of the hydrogel domain, a PhD dissertation[28] has shown that AgNPs were embedded in linoleic acid-based fibers, but the system was essentially studied using *in situ* SAXS and TEM, while the hybrid material *per se* was not stable in water. Adopting an *ex situ* reduction approach, Ikkala et al.,[29] have demonstrated that AuNPs of controlled size can be co-assembled with rod-like tobacco mosaic viruses (TMV) as a colloidally-stable suspension, demonstrating an elegant, yet complex multi-step, materials' engineering process. The gel properties of the AuNPs-TMV were not specifically reported, either, although it was mentioned that viscosity increase was proportional to the TMV content.

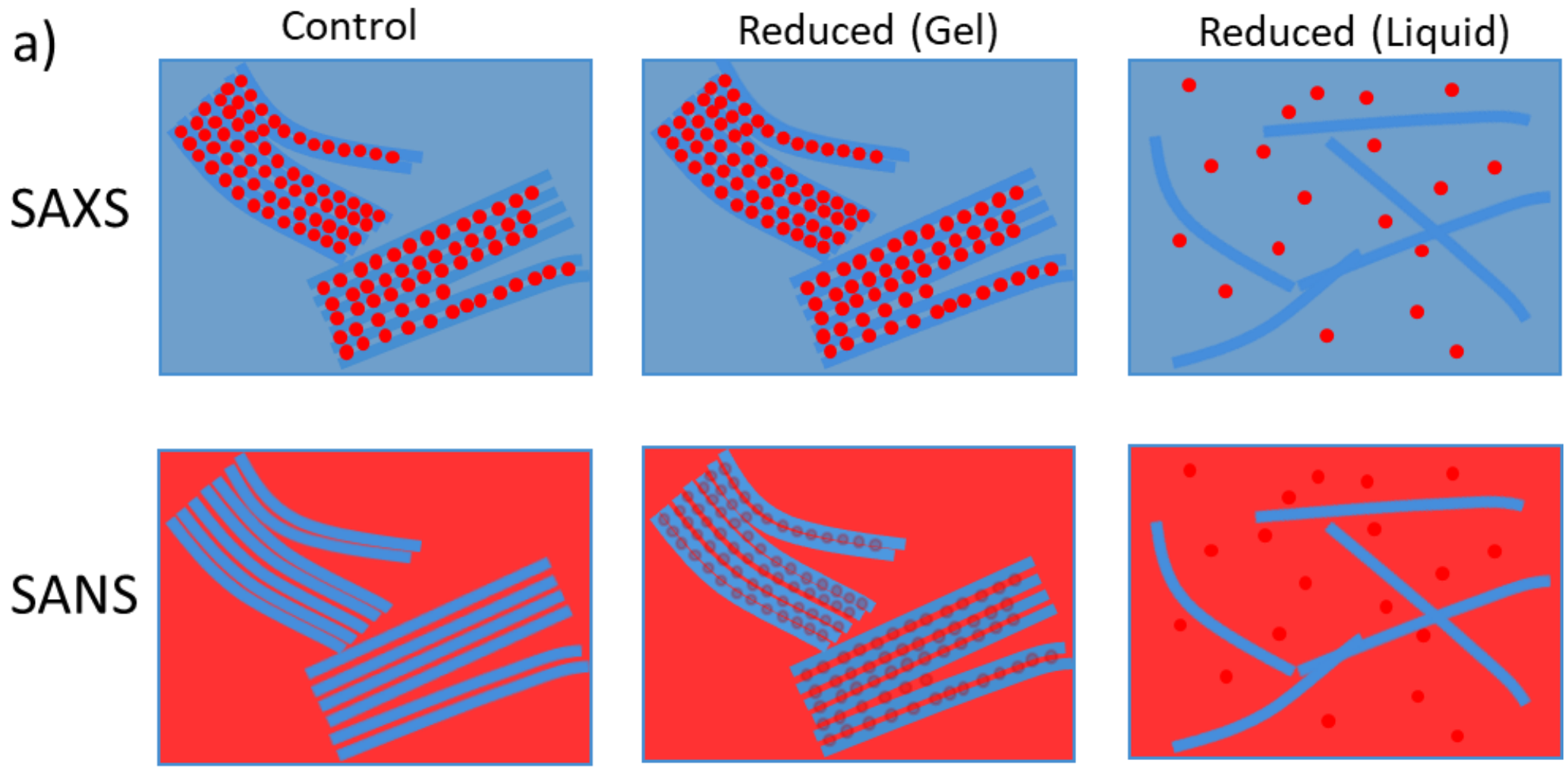

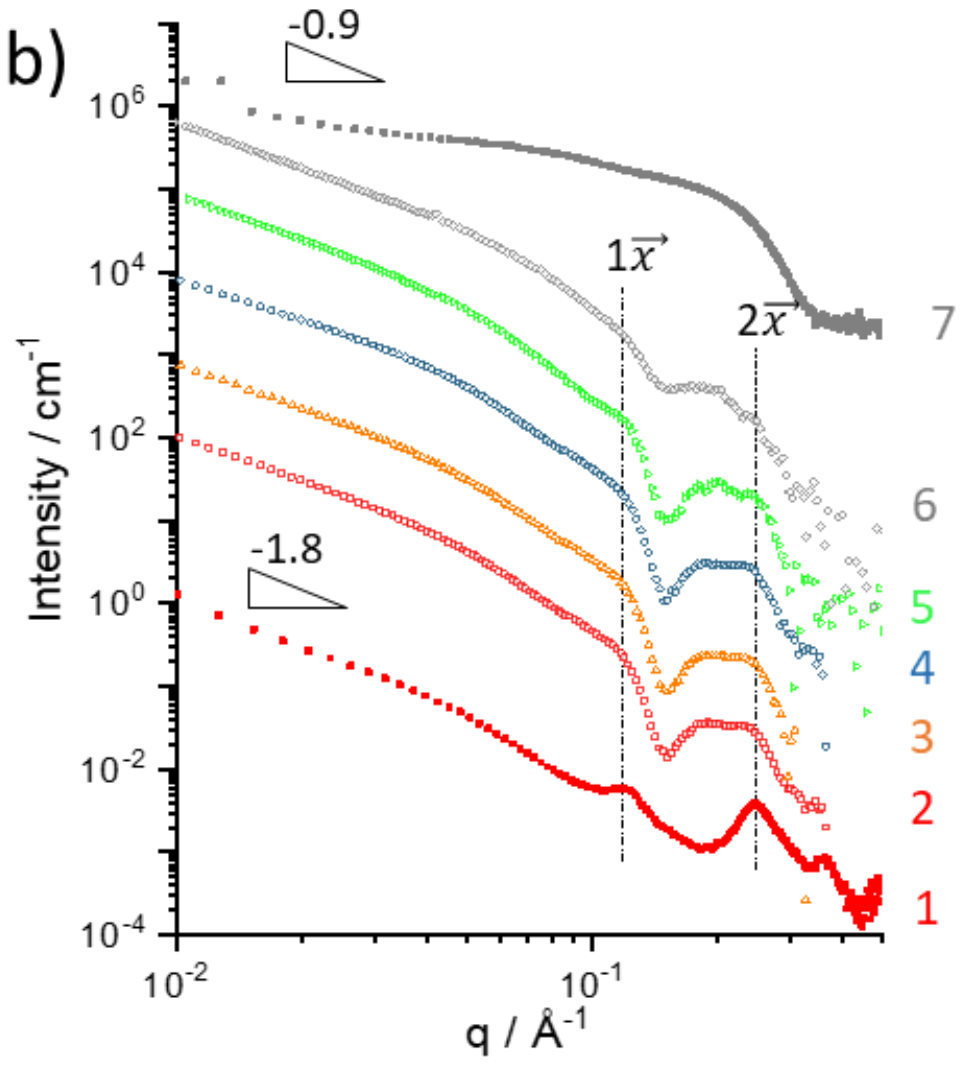

c)

| Profile | Technique | Sample |
|---|---|---|
| 7 | SAXS | *γ-rays 28h (Liquid)* |
| 6 | SANS | *γ-rays 28h (Liquid)* |
| 5 | SANS | *γ-rays 4h30 (Gel)* |
| 4 | SANS | *Ascorbic Acid (Gel)* |
| 3 | SANS | *In situ X-rays (Gel)* |
| 2 | SANS | *Control (Gel)* |
| 1 | SAXS | *Control (Gel)* |

**Figure 5- a) Cartoon explaining the contrast variation between SAXS and SANS experiments in {$Ag^+$}G-C18:1 (control) and {$Ag^0$}G-C18:1 (reduced) samples. b) SAXS and SANS profiles of c) {$Ag^+$}G-C18:1 and {$Ag^0$}G-C18:1. Profiles 1-2 and 6-7 correspond to the same sample where acquisition was made with combined SAXS/SANS.**

The system reported here generates a stable hybrid O/I gel in a pseudo-single step. Upon reduction of silver-coordinated glycolipid gels, either by the electron generated from the ascorbate free radical[79] or, by the reducing radicals generated from the radiolysis of water, $Ag^+$ ions are reduced down to $Ag^0$ atoms, which subsequently coalesce into metallic clusters, forming Ag nanoparticles. These nanoparticles further aggregate with neighbouring clusters. In the present system, the coalescence is limited by the presence of G-C18:1 molecules, whose functional groups anchor to the cluster surface, limiting the size of the clusters and maintaining a uniform particle size[80]. While the stabilization capacity of glycolipids[27] and fatty acids[28] is well known, the particularity of G-C18:1 could be attributed to its double amphiphilic properties: the fatty acid tail is able to stabilize nanoparticles, while the glucose headgroup can be hydrated, altogether driving the gel formation[43].

*From anisotropically 2D organized nanoparticles to nanowires and 3D conductive gels*

The *in situ* reduction of silver generating small, monodisperse and 2D organized (d= 2.8 nm ± 0.13%) nanoparticles raises the question of possible electron conductivity of the reduced {$Ag^0$}G-C18:1 hydrogels in the macroscopic scale. According to cryo-TEM data, tunnelling effects could occur and play a non-negligible role in the conductivity throughout the volume of the gels. Conductivity measurements of the {$Ag^0$}G-C18:1 hydrogels are carried out by impedance spectroscopy, a technique which allows to distinguish between the ionic and electronic contributions to conductivity[81]. This distinction is seldom, if never, made in the field of conductive hybrid gels, for which conductivity is measured in a standard way by applying a potential difference. In general, there is no distinction of the nature of the charge carriers[82,83] or experiments are measured at frequencies not low enough to isolate the electronic contribution[84,85].

The impedance data were analyzed using a Nyquist plot, and the results are presented in the supplementary figures (Figure S 7). A well-defined semicircle appears as the concentration of ascorbic acid increases, which is likely associated with the overlap of two conduction phenomena: ionic and electronic transport. As the concentration of ascorbic acid increases, the amplitude of the semicircle decreases, which correlates with a change in the overall electrical conductivity of the gel. Notably, the semicircle is offset, indicating the concurrent influence of both conduction mechanisms, which are challenging to decouple in this representation. For this reason, we chose to represent the data using a Bode plot, where the low-frequency plateau serves as a signature of an electronic conduction mechanism. While an equivalent circuit model could be employed to describe the impedance data, additional experiments are required to define it properly. However, such an in-depth analysis is beyond the scope of the current study.

Figure 6.a presents the classical Bode representation, where the complex impedance is plotted against frequency: the ionic contribution to conductivity is obtained from the plateau at high frequencies (between 0.1 kHz and 100 kHz) while the electronic contribution is obtained at lower frequencies (between $10^{-3}$ Hz and 100 Hz). In the ionic regime, both reduction methods display a drop of the complex impedance, explained by silver, protic ($H^+$) or ion (e.g., ascorbate, $OH^-$, $Na^+$) diffusion throughout the gel, increasing the ionic conduction.[86] At lower frequencies, on the contrary, both the control and samples reduced by radiolysis display a monotonous increase of the modulus of the impedance, an evidence of no electrical conductivity. On the other hand, a plateau is reached below $10^{-1}$ Hz for the sample reduced with ascorbic acid, indicating electronic contribution to conductivity.

To better understand the possible conductive mechanism in {$Ag^0$}G-C18:1-AA, impedance spectroscopy, cryo-TEM and SAXS are combined on a set of samples prepared by varying the concentration of ascorbic acid added to the {$Ag^+$}G-C18:1 gels, from 5 µL/mL to 80 µL/mL (Figure 6b-f). By controlling the content of ascorbic acid, an additional structure, identified as nanowires by cryo-TEM (Figure 6.c,e) is observed. The chemical composition combined to the FFT analysis performed on the HRTEM images shown in Figure S 2 confirms the silver metal nature of the wires.

The increasing AA content also induces the appearance of a new peak (noted $q_{iw}$, iw for inter-wire) in the range between 0.04 Å$^{-1}$ and 0.06 Å$^{-1}$, as shown by the Kratky plot in Figure 6.f. The origin of this peak, characteristics of larger distances, could be attributed to the repeating inter-wire distance ($d_{iw}$), in well agreement with a diffraction spot at d= 114.8 Å ($\boldsymbol{q} = \frac{\boldsymbol{2\pi}}{\boldsymbol{d}}$= 0.054 Å$^{-1}$ ) observed on the FFT analysis performed on cryo-TEM images (Figure 6.c,d). A comparative analysis between cryo-TEM and SAXS is also provided in Figure S 8. The $q_{iw}$ from SAXS (q= 0.04 - 0.06 Å$^{-1}$) and $d_{iw}$ from cryo-TEM (q= 0.054 Å$^{-1}$) identify the same plane. The actual origin of the fibers could be explained by the AA-induced fusion of the nanoparticles (Figure 3, Figure 4) along $\vec{\boldsymbol{z}}$ and possibly $\vec{\boldsymbol{x}}$.

Such a mechanism has been previously observed by Giersig et al.[87], where Ag nanowires formed at relatively high silver salt concentration upon aggregation of neighbouring nanoparticles through the alignment of the same adjacent crystallographic planes. Another example has been observed in Pt-doped mesophases where 1D and 2D-assemblies of Pt NPs were obtained respectively in hexagonal and lamellar mesophases, by slow reduction of Pt complexes by CO[88] and Krishnaswamy et al.[89] showed that $Pt^0$ nanoparticles obtained by radiolysis in lamellar mesophases could grown into Pt nanorods in time, by coalescence of small Pt seeds. Similarly, in this study, the possible growth of Ag nanowires over time is put in evidence by the increased prominence of the peak associated with the inter-wire distance $q_{iw}$ (length scale of one hundred nm), after 2 weeks, as illustrated in the Kratky plot shown in Figure S 9a. However, this maturation over time didn't necessarily lead to an increase in the conductive properties (Figure S 9b), probably due to the loss of the three-dimensional (ten of microns length scale) organization of the wires.

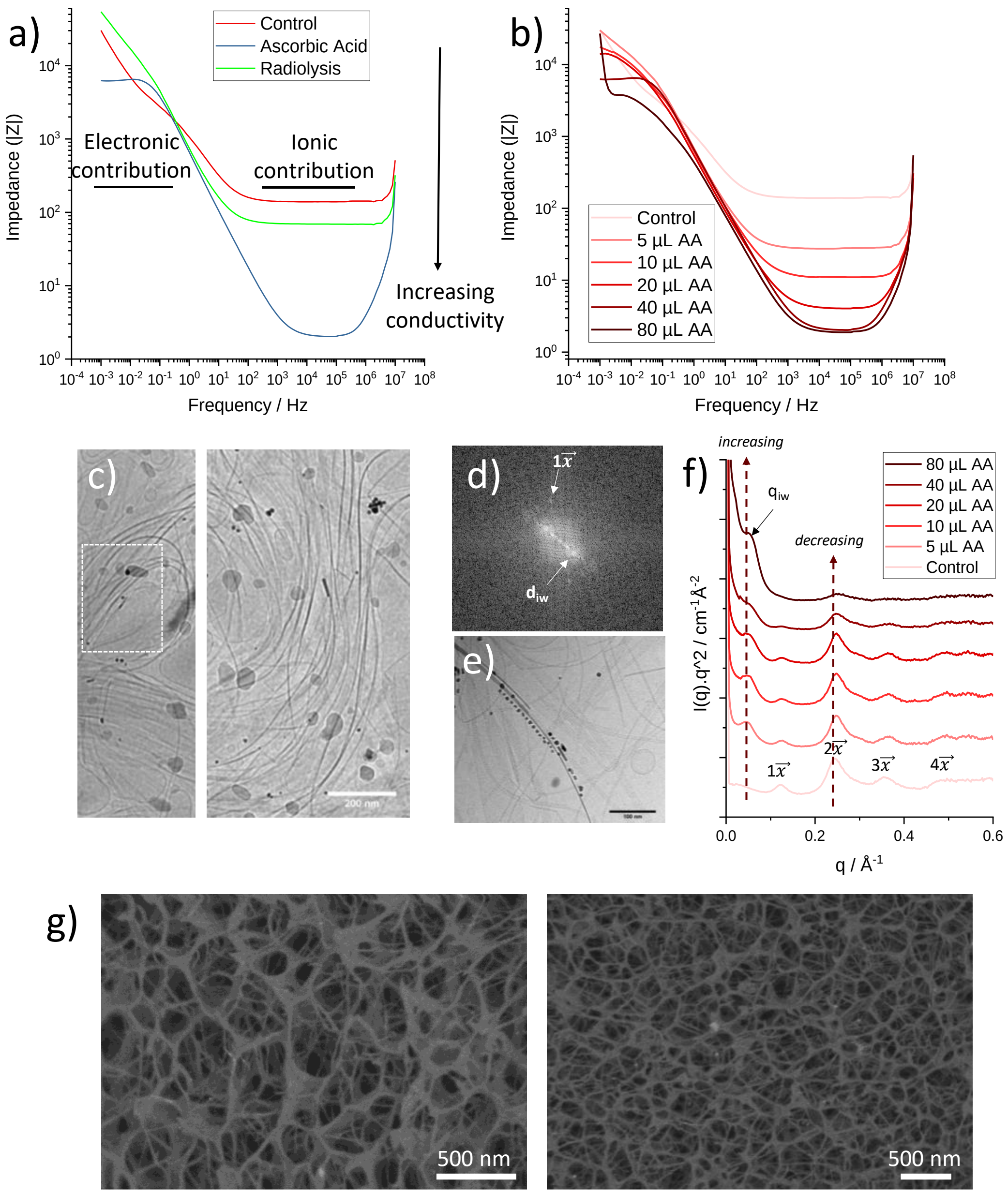

**Figure 6. Impedance spectroscopies (IS) of a) non-reduced $\{Ag^+\}$G-C18:1 gel (control) and $\{Ag^0\}$G-C18:1 gels reduced by ascorbic acid and radiolysis and b) IS of $\{Ag^0\}$G-C18:1-AA gels with increasing ascorbic acid (AA) content from 5 to 80 µL/mL. Impedance is inversely related with conductivity and low frequencies represent the electronic contribution while high frequencies demonstrate the ionic contribution. c) Cryo-TEM image of $\{Ag^0\}$G-C18:1-AA gel, 2 weeks after reduction, and d) FFT pattern (corresponding to the white rectangle in c), with $d_{iw}$ for inter-wire distance. e) Magnified Cryo-TEM images of the wires. f) Scaled Kratky plots of samples with increasing AA content: a new peak at $q_{iw}$ corresponding to $d_{iw}$, appears and is shown with an arrow while the intensity of the peaks $1\vec{x}$, $2\vec{x}$, $3\vec{x}$, $4\vec{x}$ decreases. g) Cryo-SEM images of the $\{Ag^0\}$G-C18:1-AA gels (AA content of 40 µL/mL).**

In the current work, this hypothesis is corroborated by the gradual loss in intensity of the $1\vec{x}$, $2\vec{x}$, $3\vec{x}$, $4\vec{x}$ and $1\vec{z}$ Bragg peaks with increasing AA content (Figure 6.f), recalling from the SAXS-SANS data (Figure 5.b) that these reflections are associated to the contrast between silver NPs and water. At the highest content of AA probed here (80 µL/mL), the peak at $q < 0.1$ $Å^{-1}$ reaches the highest intensity, while all peaks above 0.1 $Å^{-1}$ have almost disappeared. Although the kinetics of wire formation is not studied in detail, we must alert on the fact that only the gels prepared with AA content below 20 µL/mL are stable in time (months). The gels prepared with AA content above this threshold undergo syneresis and eventually precipitation within hours/days. One can speculate that the loss in the elastic properties in time is correlated to the loss of a three-dimensional network through a bundling mechanism, for instance.

The impedance spectroscopy diagrams (Figure 6.b) corresponding to the SAXS data show that the addition of ascorbic acid is correlated to a plateau of lower and lower complex impedance in the electronic conduction regime. Adding between 5 and 20 µL/mL of AA reduces the slope of the modulus of impedance below 0.1 Hz until a pseudo-plateau below $10^{-3}$ Hz (the lowest experimental limit) at 20 µL/mL. Above 40 µL/mL, the plateau becomes much better defined and of lower impedance already below 0.1 Hz, thus demonstrating an increase in the electrical component of the conductivity. For the sample at 80 µL/mL, one observes a drastic increase in the impedance below $2 \cdot 10^{-3}$ Hz, a phenomenon, which could be explained by the long acquisition time (1h30), considering that syneresis occurs at high AA content. In other words, the loss in the three-dimensional interconnectivity of the gel could be responsible for the loss in the electrical conductivity of the measuring cell.

Notably, the tendency of the wires to assemble into macroscopic 3D structures is also the reason why $\{Ag^0\}$G-C18:1-AA samples pose challenges for cryo-TEM imaging and rheo-impedance spectroscopy. As a result, cryo-TEM might be inadequate to capture a representative view of the macroscopic wire network. While this technique can be used to prove the presence of wires in 2D scale, it cannot provide a full understanding of their 3D organization. Therefore, additional cryo-SEM experiments, shown in Figure 6.g, help better visualizing the extensiveness of the three-dimensional wire network at the micron scale. The limited spatial resolution of SEM is backed up in Figure 7 by complementary TEM and STEM experiments, which confirm presence, continuity and "infinite" length of the nanowires, shown by the black arrows (Figure 7.a), as well as their silver composition (Figure 7.b). Such unique combination of super- and sub-micron analysis allows a better understanding about both the microscopic and macroscopic properties of this multi-scale material. One should note that systems reduced by AA may also display regions of high heterogeneity, where both individual fibers and bundles coexist with silver nanoparticles (Figure 7.a, Figure S 10). It cannot be excluded that such a heterogeneity, probably due to unoptimized reduction conditions, could have a deleterious impact on the conductivity performances. Further work is needed to improve and optimize the structure-properties relationship of {$Ag^0$}G-C18:1 metallogels.

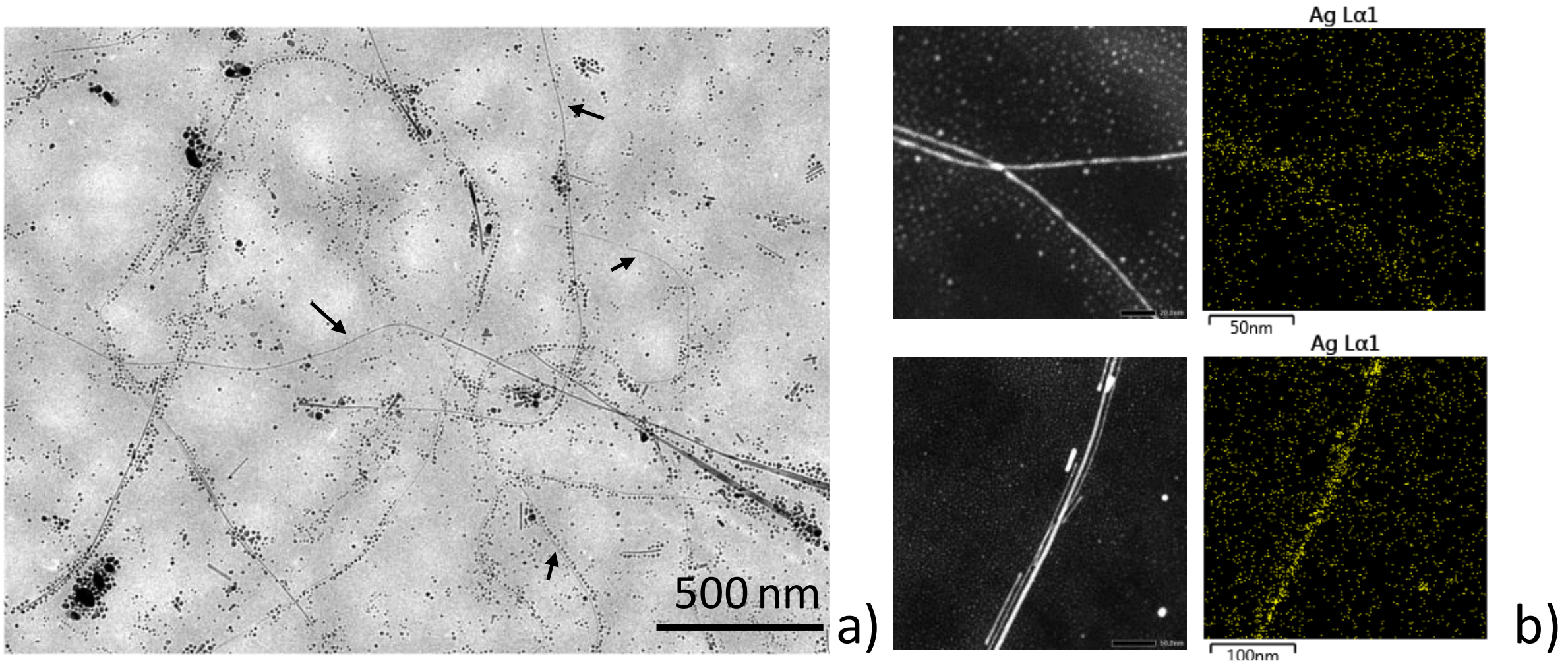

**Figure 7. a) TEM image of {$Ag^0$}G-C18:1-AA gel (diluted 20 times) 2 weeks after reduction and b) selected STEM images with corresponding Ag elemental analysis.**

## Conclusions

In this work, we present the *in-situ* formation of silver nanoparticles and nanowires aligned in the fibers of the metallogels of G-C18:1, a microbial glycolipid, by different reduction methods. Among the reducing methods applied, the use of sodium borohydride ($NaBH_4$) as the strongest reductant, destroyed the gel completely, while radiolysis by γ-rays, the weakest reduction method, led to the formation of homogeneous silver nanoparticles along the gel fibers. In the case of ascorbic acid, a mild reductant, the formation of silver nanowires, in addition to nanoparticles, is observed, keeping the gel structure intact. The gels containing silver nanoparticles and nanowires are characterized by rheology and elastic properties similar to the as-prepared metallogel are observed. Cryo-TEM images show the synthesis of monodispersed Ag nanoparticles aligned with the fibers upon reduction by radiolysis and ascorbic acid, with an additional formation of nanowires with the latter, and all by keeping the raft-like association of the fibers unaffected. SAXS measurement agrees with the images, where the gels have similar profiles, either reduced or as-prepared, and a peak appears for the sample with ascorbic acid, likely corresponding to the inter-fibrillar distance emerging due to the formation of silver nanowires. UV-Vis spectroscopy gives evidence of the formation of Ag nanomaterials (nanoparticles or nanowires) with the appearance of characteristic peaks upon $Ag^+$ reduction. However, FTIR results indicate that the complexation between the $COO^-$ end of G-C18:1 and silver remains the same whether it is $Ag^+$ or $Ag^0$. Silver nanowire-embedded gels demonstrate an electronically conductive behavior according to impedance spectroscopy and the conductive property is observed to increase with the amount of nanowires formed in the gel, which might be explained by the extensiveness of the three-dimensional wire network on the micron scale, as observed via cryo-SEM experiments. The impedance measurement which is employed here can differentiate the ionic conductivity due to diffusion and migration of ions from the electronic conductivity arising from electron transfer. This differentiation is achieved through measurements in a wide frequency range. Unlike discussions from much of the existing literature, focusing on a mixed conductivity value, here we present electrical conductivity and ionic conductivity separately. The silver nanoparticle-embedded gels do not exhibit electronic conductivity, however the exceptional alignment of uniform metal nanoparticles in gel systems, achieved here through the unique properties of G-C18:1 when compared to other fatty acids[28], has not been previously reported[9]. The remarkable dual role of G-C18:1 to self-assemble into fibrillar structures and stabilizing the nanoparticles, allowing a single-component material to address multiple challenges within one system, distinguishes this work from existing literature[29]. One should note that while the 2D properties of these nanostructures have been explored by UV-Vis, cryo-TEM, SAXS and SANS, the key strength of this study is the understanding of the 3D organization of the structures observed through macroscopic conductivity, by means of impedance spectroscopy and cryo-SEM.

**Acknowledgments**

ILL, Grenoble, France, is kindly acknowledged for financial support during the EASY-1216 proposal. We acknowledge support from the Partnership for Soft Condensed Matter (PSCM) at the ILL. The French Microscopie Electronique en Transmission et Sonde Atomique (METSA) network is kindly acknowledged for providing financial support on the 200 kV cryo-TEM microscope. Jeril Degrouard and Amélie Le Forestier (Laboratoire de Physique des Solides, Orsay, France) are kindly acknowledged for their support in the cryo-TEM experiments. Michel Goldmann (Institut des Nanosciences de Paris, Sorbonne Université, France) et Marianne Impéror-Clerc (Laboratoire de Physique des Solides, Orsay, France) are acknowledged for helpful discussions.

# Supplementary Information

## Electron conductive self-assembled hybrid low-molecular weight glycolipid-nanosilver gels

Korin Gasia Ozkaya[a], Othmane Darouich,[a] Hynd Remita[b], Isabelle Lampre[b], Lionel Porcar[c], Alain Carvalho[d], Marc Schmutz[d], Sandra Casale,[e] Christel Laberty-Robert[a], Niki Baccile[a,*]

[a] *Sorbonne Université, Centre National de la Recherche Scientifique, Laboratoire de Chimie de la Matière Condensée de Paris (LCMCP), UMR 7574, F-75005 Paris, France*

[b] *Université Paris-Saclay, CNRS, Institut de Chimie Physique (ICP), UMR 8000, Faculté des Sciences d'Orsay, 91405 Orsay, France*

[c] *Institut Laue Langevin, 38042 Grenoble, France*

[d] *Université de Strasbourg, CNRS, Institut Charles Sadron UPR 22, 67034 Strasbourg, France*

[e] *Sorbonne Université, Centre National de la Recherche Scientifique, Laboratoire de Réactivité de Surface (LRS), UMR 7574, F-75005 Paris, France*

**Niki Baccile,* niki.baccile@sorbonne-universite.fr

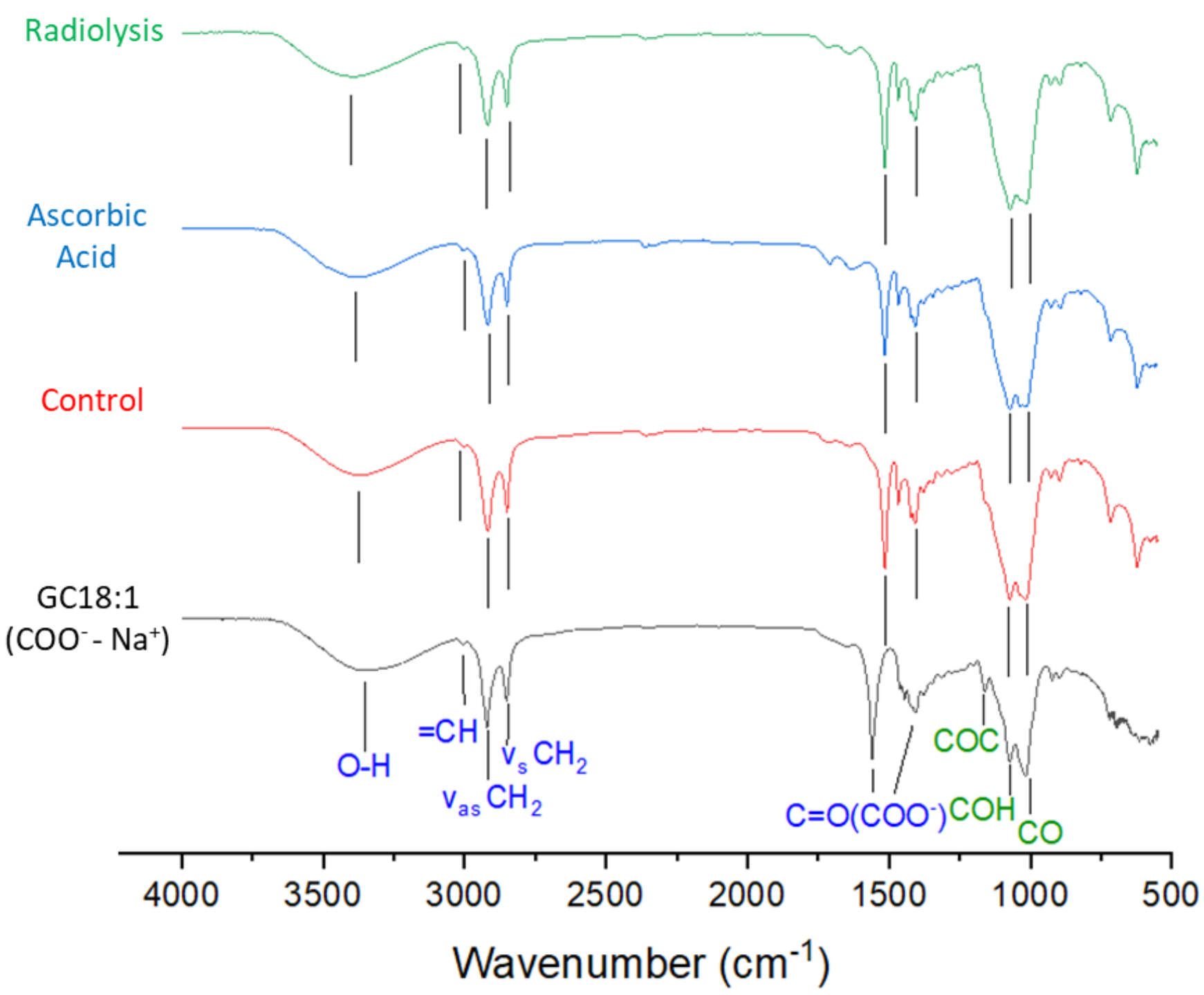

**Figure S 1. Full FTIR spectra on freeze-dried samples.**

Blue annotated peaks belonging to the aliphatic backbone and carboxylate group remained unchanged upon gel formation and reduction. The broad peak observed at 3344 $cm^{-1}$ corresponds to O-H stretch and the small peak at 3005 $cm^{-1}$ corresponds to CH groups of the alkene. The bands at 2921 $cm^{-1}$ and 2852 $cm^{-1}$ are due to asymmetrical stretching ($v_{as}$ $CH_2$) and symmetrical stretching ($v_s$ $CH_2$) of methylene, respectively[1]. Similarly, the green annotated bands at 1161 $cm^{-1}$, 1075 $cm^{-1}$ and 1015 $cm^{-1}$, corresponding to the ether or alcohol linkages found in the glucose groups, stay intact upon gelification and further reduction as well. This shows that the structure of G-C18:1 doesn't change after reduction.

The spectrum of GC18:1 at pH 8, presents bands at 1560 $cm^{-1}$ and 1406 $cm^{-1}$, corresponding to $COO^-$ asymmetrical and symmetrical stretching, when the carboxylate group is coordinated with $Na^+$ ion. Upon gel formation due to $COO^-$-$Ag^+$ coordination, the band due to $v_{as}$ $COO^-$ at 1560 $cm^{-1}$ shifts to 1510 $cm^{-1}$, and the frequency difference between $COO^-$ antisymmetric and symmetric vibrations, $\Delta v(=v_{as}-v_s)$, changes from $\Delta v$=154 to $\Delta v$=104. This frequency difference can be explained by the change in the coordination type of the carboxylate group from bridging bidentate with $Na^+$ ions, to cheating bidentate with $Ag^+$ ions[2]. Upon reduction by ascorbic acid or radiolysis, these bands stay intact, showing that the coordination of $COO^-$ with $Ag^+$ cations or $Ag^0$ nanoparticles remains identical.

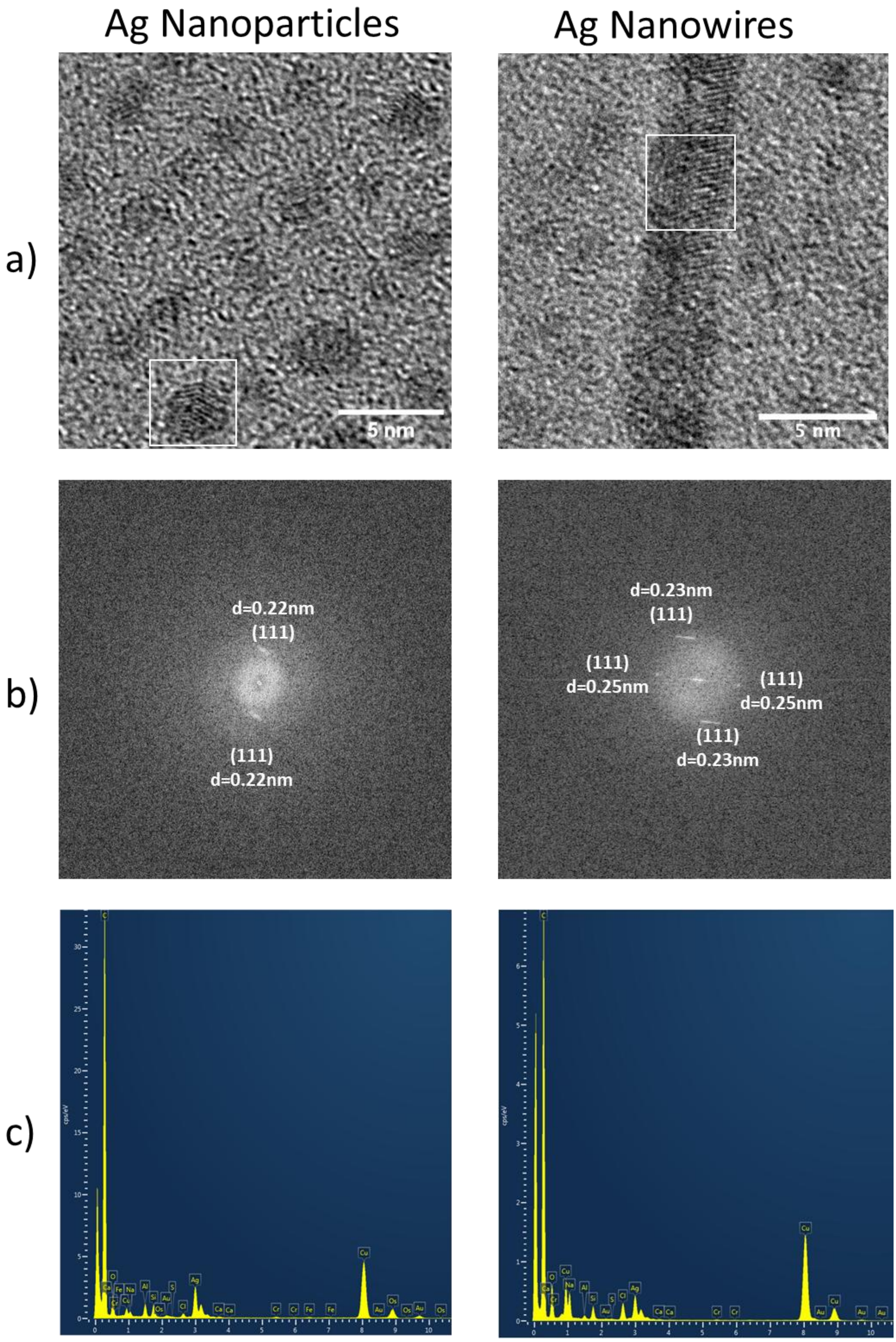

**Figure S 2. a) High Resolution TEM images of nanoparticles and nanowires obtained upon reduction and b) the Fast Fourier Transform (FFT) patterns corresponding to the white squared regions in a). c) Spectra of EDX chemical analysis.**

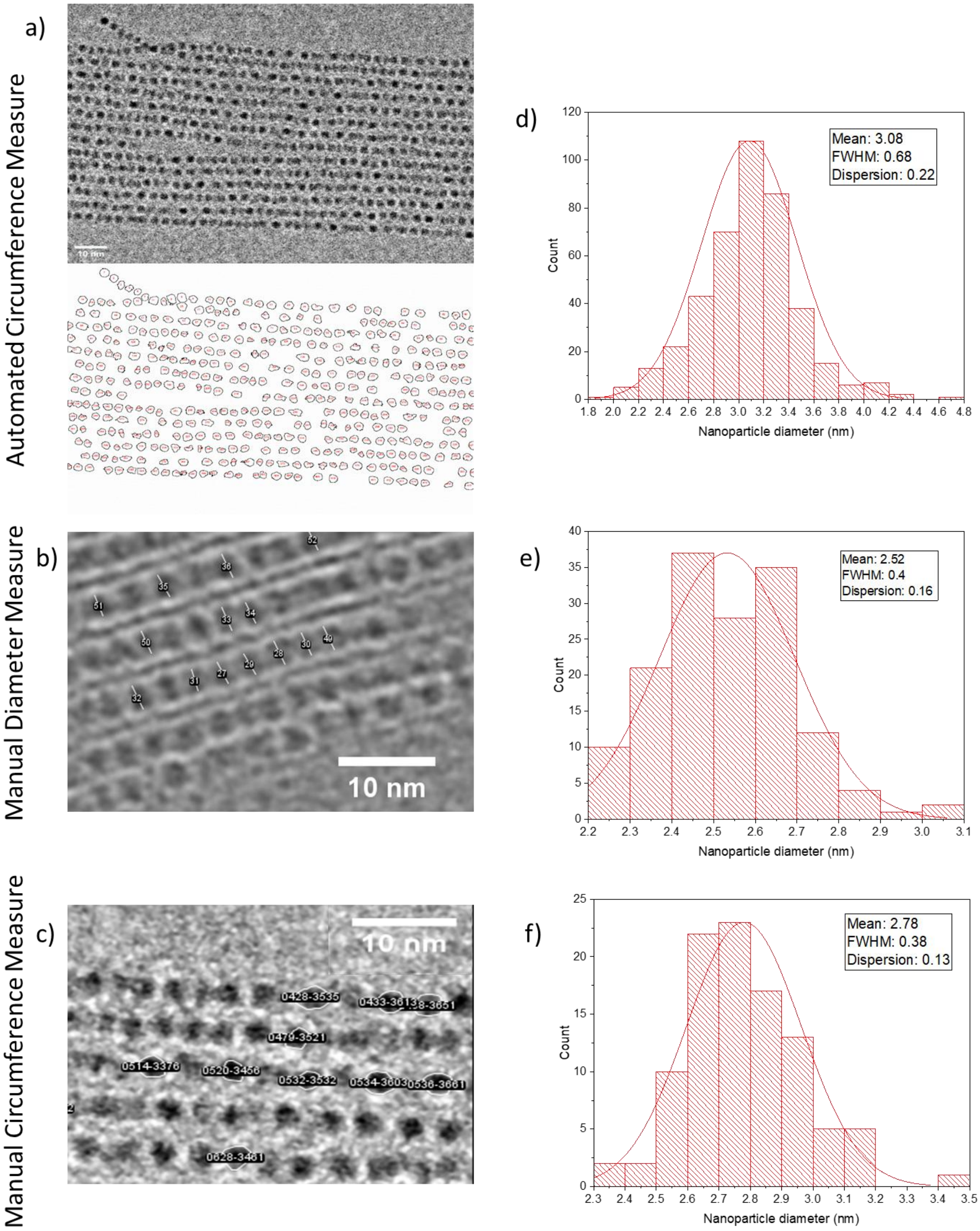

**Figure S 3. Evaluation of the size of Ag nanoparticles visualized by cryo-TEM images. a) Automated method to measure the circumference of the particles by an ImageJ plug-in called "Trainable Weka Segmentation"; b) manual measurement of the diameter of the particles; c) manual measurement of the circumference of the particles and d,e,f) the corresponding size distribution profiles for each method.**

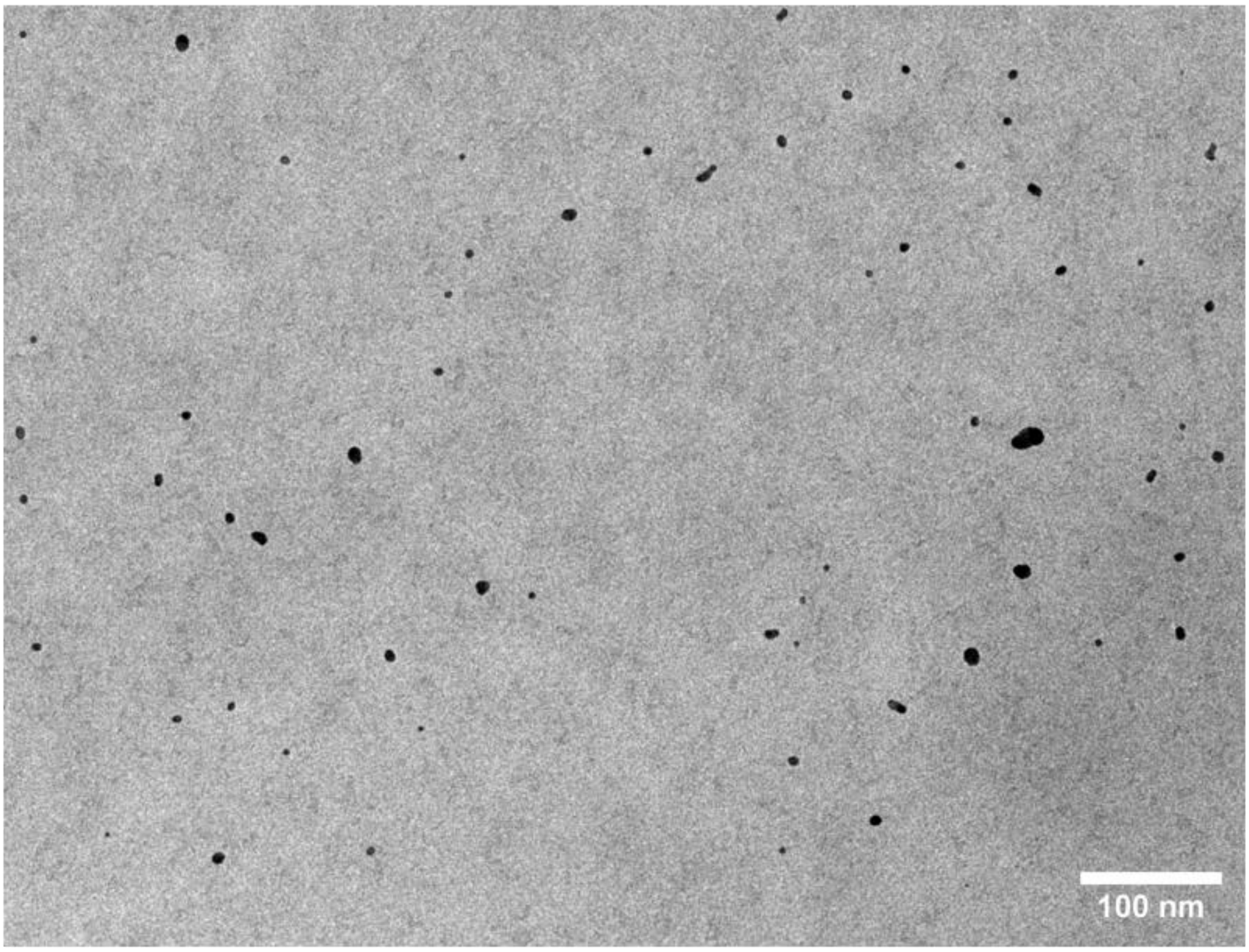

**Figure S 4. Cryo-TEM image of {$Ag^0$}G-C18:1-$NaBH_4$ sample.**

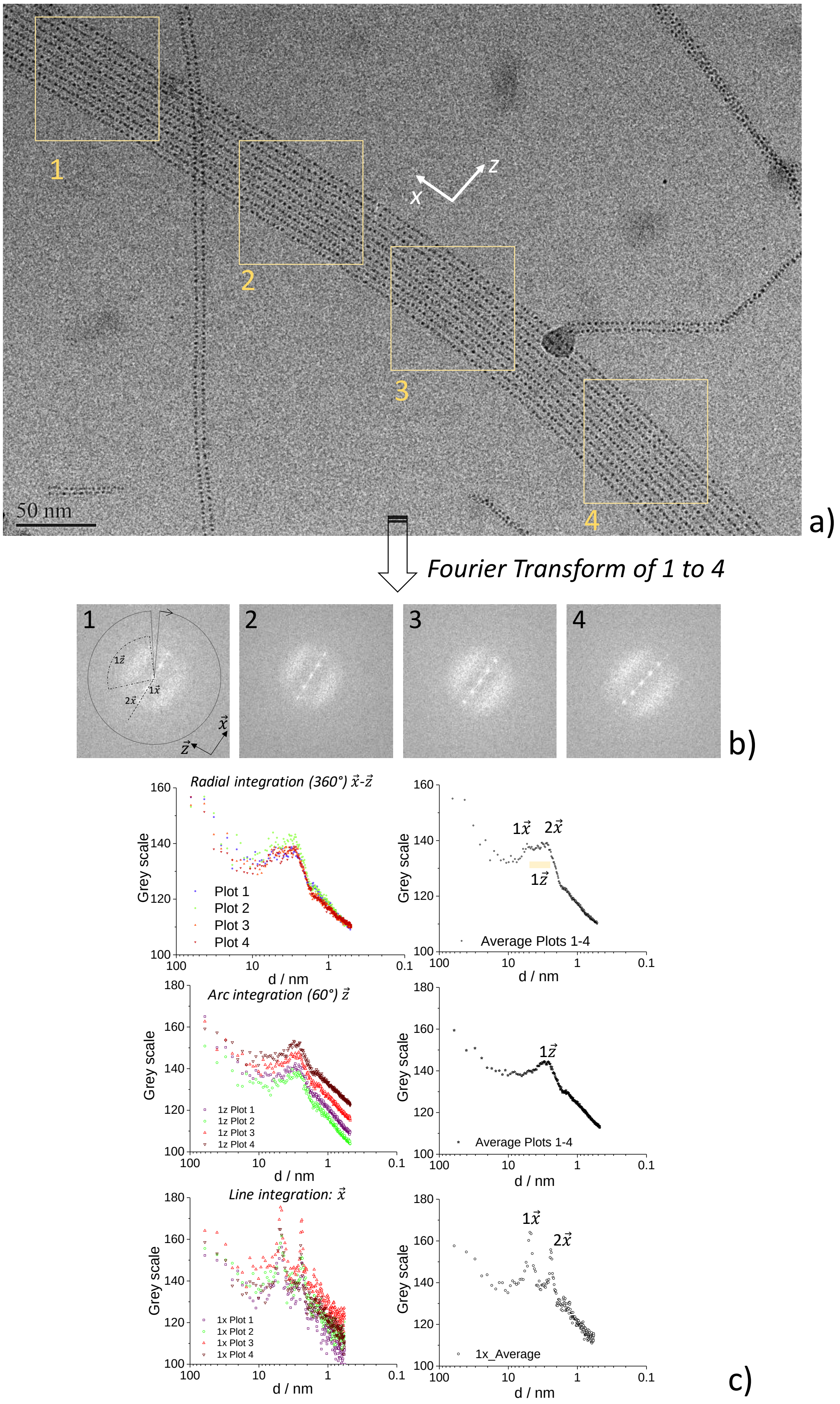

**Figure S 5. a) Typical cryo-TEM image showing the organization of Ag nanoparticles upon reduction; b) FFT patterns of 4 regions selected within yellow squares in a); c) radial (360°), arc (60°) and line integrations done on FFT patterns of the 4 different regions, and their averages where $1\vec{x}$, $2\vec{x}$ and $1\vec{z}$ peaks are labeled.**

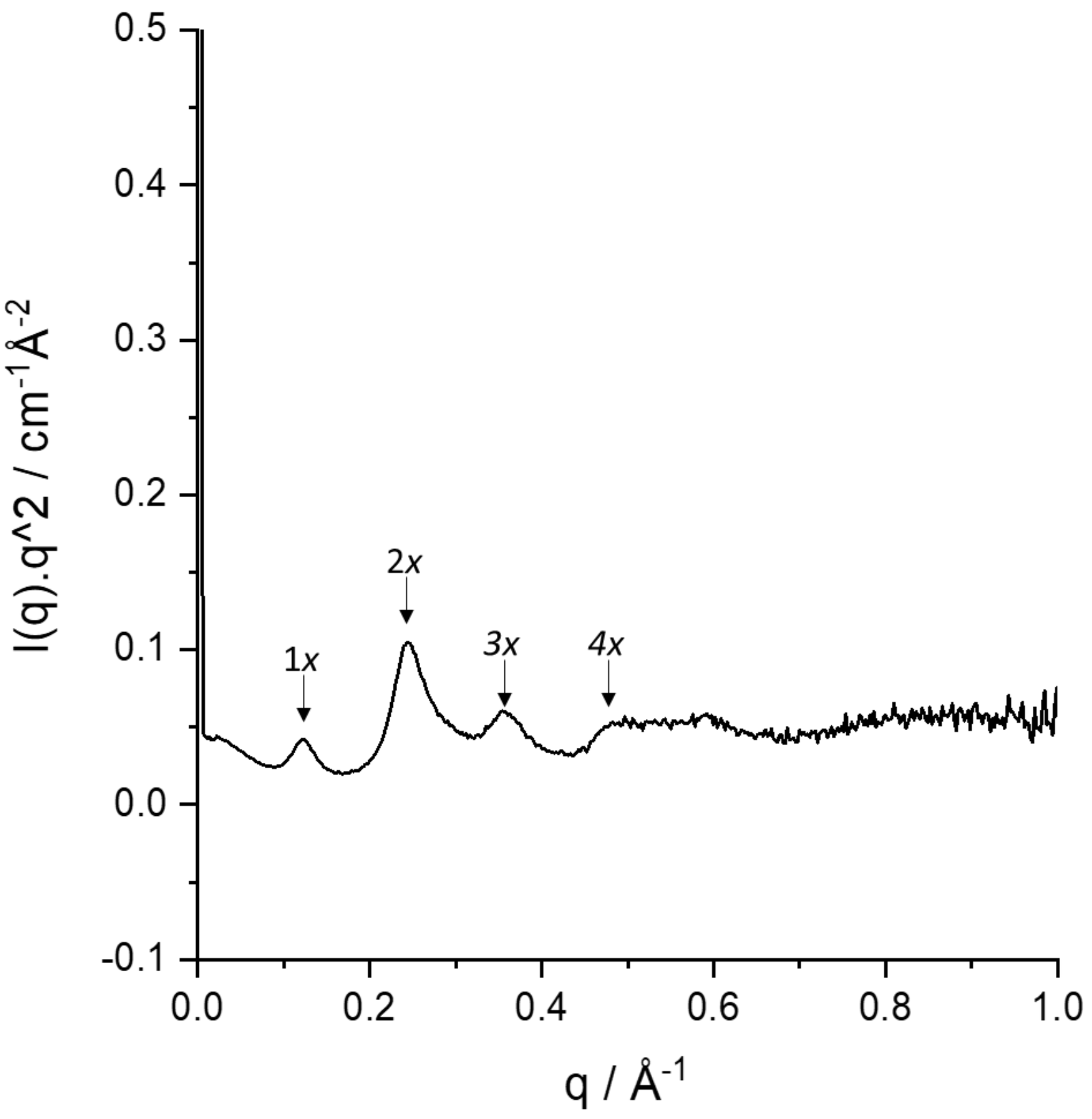

**Figure S 6. Kratky plot of a SAXS measurement done for {$Ag^{+}$}G-C18:1.**

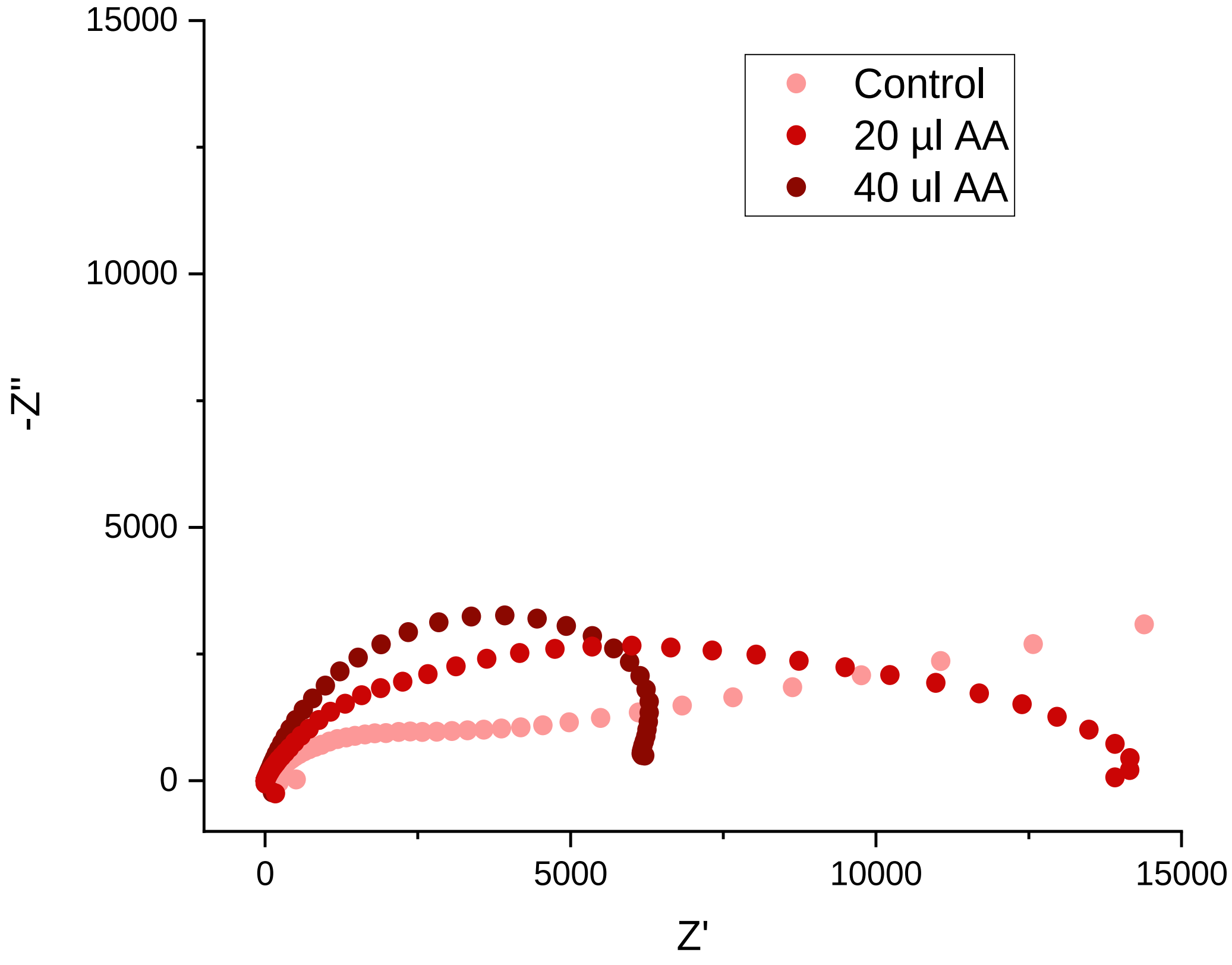

**Figure S 7. Nyquist plots of {$Ag^{+}$}G-C18:1 control and {$Ag^{0}$}G-C18:1 samples prepared with two different content of ascorbic acid (AA)**

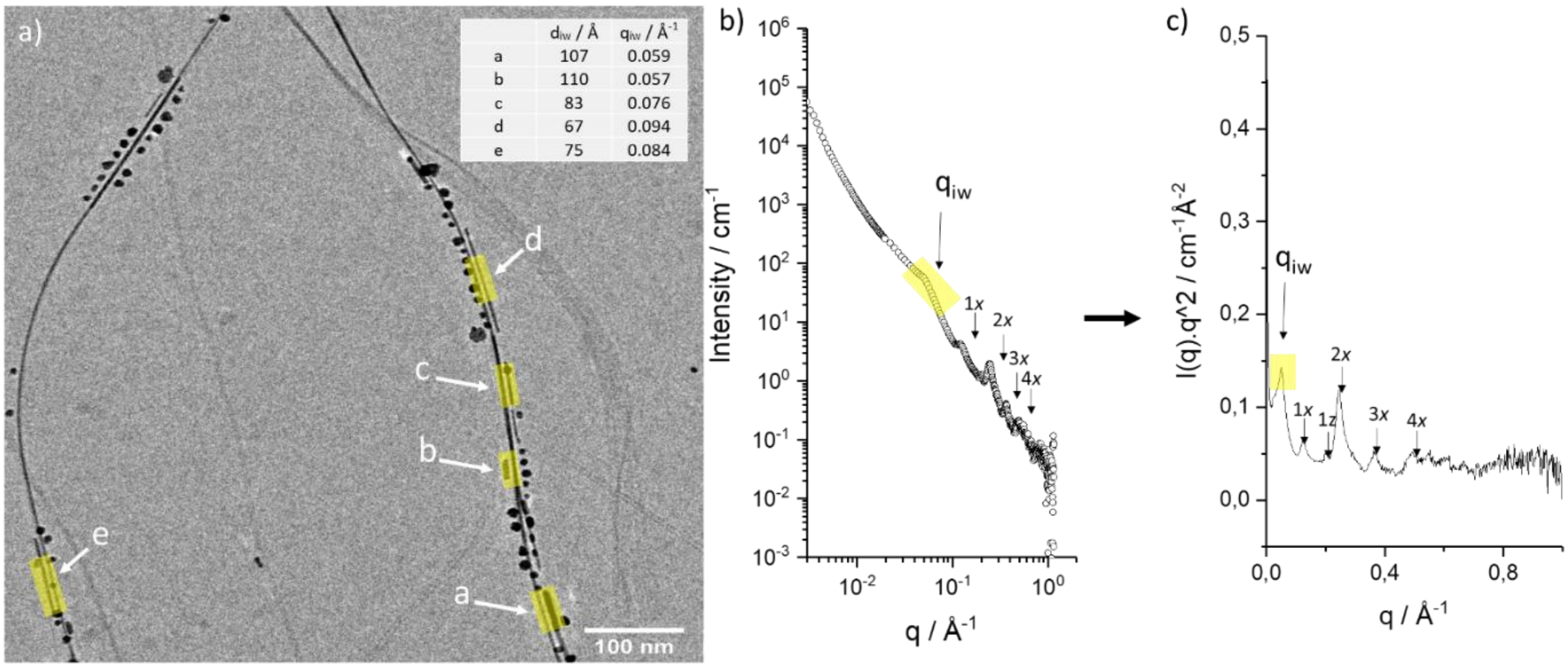

**Figure S 8. a) Cryo-TEM image of silver nanowires formed upon reduction by ascorbic acid. The inter-wire distance ($d_{iw}$) is calculated manually from the yellow areas and noted in the table, with the corresponding q values. b) SAXS measurement for the same sample, with c) the corresponding Kratky plot. The peak highlighted with yellow has a q value ($q_{iw}$) coherent with the measured interwire values.**

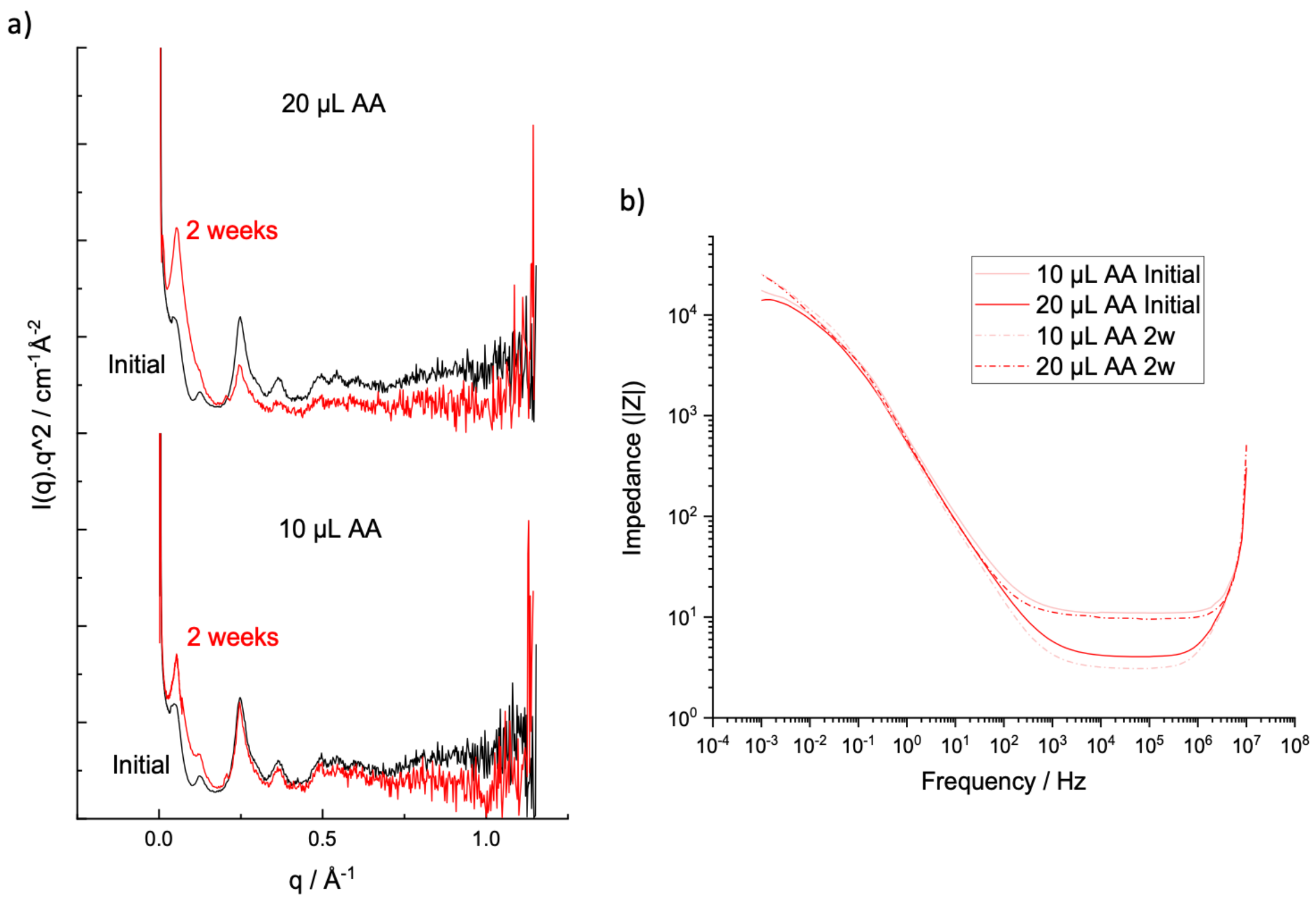

**Figure S 9. a) Kratky plots of the SAXS measurements and b) impedance spectroscopy measurements for {$Ag^0$}G-C18:1-AA gels (AA content of 10 and 20 µL/mL) performed initially and after 2 weeks.**

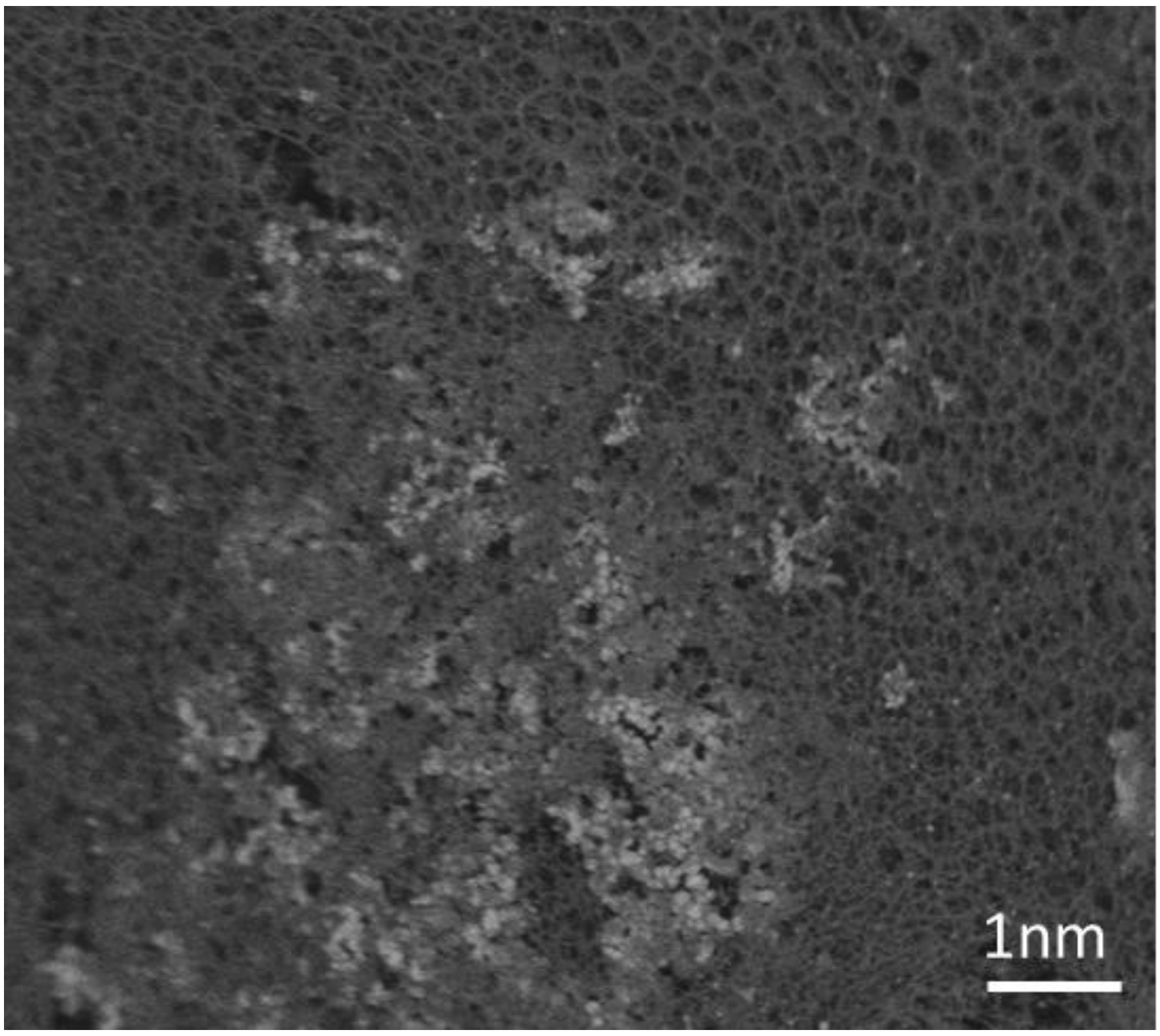

**Figure S 10. Cryo-SEM image of a {Ag$^{0}$}G-C18:1-AA gel (AA content of 40 µL/mL) where coexistence of the 3D fiber network and nanoparticles can be observed.**